\documentclass[showpacs,twocolumn,superscriptaddress]{revtex4}

\usepackage{graphicx}
\usepackage{dcolumn}
\usepackage{amsmath}

\begin{document}
\title{Shapes of rotating nonsingular black hole shadows}

\author{Muhammed Amir}
\email{amirctp12@gmail.com}

\affiliation{Centre for Theoretical Physics,
 Jamia Millia Islamia,  New Delhi 110025,
 India}

\author{Sushant G. Ghosh}
\email{sghosh2@jmi.ac.in}

\affiliation{Centre for Theoretical Physics,
 Jamia Millia Islamia,  New Delhi 110025,
 India}

\affiliation{Astrophysics and Cosmology Research Unit,
 School of Mathematics, Statistics and Computer Science,
 University of KwaZulu-Natal, Private Bag X54001,
 Durban 4000, South Africa}

\begin{abstract}
It is believed that curvature singularities are a creation of general relativity and hence, in the absence of a quantum gravity, models of nonsingular black holes have received significant attention. We study the shadow (apparent shape), an optical appearance because of its strong gravitational field, cast by a nonsingular black hole which is characterized by three parameters, i.e., mass ($M$), spin ($a$), and a deviation parameter ($k$). The nonsingular black hole under consideration, is a generalization of the Kerr black hole {that} can be recognized asymptotically ($r>>k, k>0$) explicitly as the Kerr-Newman black hole, and in the limit $k \rightarrow 0$ as the Kerr black hole. It turns out that the shadow of a nonsingular black hole is a dark zone covered by {a} deformed circle. Interestingly, it is seen that the shadow of a black hole is affected due to the parameter $k$. Indeed, for a given $a$, the size of a shadow reduces as the parameter $k$ increases and the shadow becomes more distorted as we increase the value of the parameter $k$ when compared with the analogous Kerr black hole shadow. We also investigate, in detail, how the ergoregion of a black hole is changed due to the deviation parameter $k$.
\end{abstract}
\pacs{04.70.Bw, 04.50.-h, 04.70.-s}

\maketitle

\section{Introduction}
The determination of  black hole spin is an open issue in astrophysics that has received significant attention using various methods  \cite{Tanaka:1995en,Fabian:1995qz,Genzel:2003as,Connars:1980,Asada:2003nf,Tanaka:1996ht,Finn:2000sy}. It turns out that {the astronomical} observation of a black hole shadow (apparent shape) may significantly help us to resolve this  important open issue. It is widely believed that the black hole shadow may contain information about the nature of a black hole, and it can help us to determine the black hole spin \cite{Takahashi:2004xh}. It is {shown} that the shape of a rotating black hole is deformed in an optically thin emitting medium due to the presence of spin \cite{Bardeen:1973gb,Chandrasekhar:1992,Falcke:1999pj}, which for the Schwarzschild black hole or a nonrotating black hole is circular \cite{Bozza:2009yw}.

It is also believed that a supermassive black holes exist in the center of a galaxy. Falcke \cite{Falcke:1999pj} determined the formation of images of supermassive black holes to demonstrate that a strong gravitational field bends the trajectories of photons released by {a} bright object and an observer can see a dark zone around a black hole. In 2002, Holz and Wheeler \cite{Holz:2002uf} detected the retro-images or retro-MACHOS of the sun by a Schwarzschild black hole. In \cite{Takahashi:2004xh}, the author discussed the qualitative and quantitative analyses of the shape and position of the rotating black hole shadow on an optically thin accretion disk and its dependence on the angular momentum. 

A black hole casts a shadow as an optical appearance because of its strong gravitational field, which  for the Schwarzschild black hole is investigated in the pioneer study of Synge \cite{Synge:1966} and Luminet \cite{Luminet:1979} who also suggested a formula to calculate the angular radius of the shadow \cite{Synge:1966}. It is shown that a shadow of the Schwarzschild black hole is a perfect circle \cite{Bozza:2009yw} through the gravitational lensing either in vacuum \cite{Virbhadra:2008ws} or in plasma \cite{Morozova:2013dxb,Bisnovatyi-Kogan:2015dxa}. Bardeen \cite{Bardeen:1973gb} studied the appearance of a shadow cast by the Kerr black hole (see also \cite{Chandrasekhar:1992}). It can be seen for the Kerr black hole that the shadow is no longer circular; it has a deformed or distorted shape in the direction of rotation \cite{Bozza:2007gt,Chandrasekhar:1992,Falcke:1999pj,Zakharov:2005ek,Nedkova:2013msa,Bambi:2008jg}. The shadow of the Kerr black hole or a Kerr naked singularity by constructing observables has been discussed by Hioki and Maeda \cite{Hioki:2009na}. The shadows of several black holes and naked singularities, have been rigorously studied in the last few years, e.g., for the Kerr-Newman black hole \cite{Takahashi:2005hy}, Einstein-Maxwell-Dilaton-Axion black hole \cite{Wei:2013kza}, Kerr-Taub-NUT black hole \cite{Abdujabbarov:2012bn}, rotating braneworld black hole \cite{Amarilla:2011fx}, Kaluza-Klein rotating dilaton black hole \cite{Amarilla:2013sj}, rotating non-Kerr black hole \cite{Atamurotov:2013sca}, Kerr-Newman-NUT black holes with cosmological constant \cite{Grenzebach:2014fha}, and the Kerr black hole with scalar hair \cite{Cunha:2015yba}. A coordinate-independent characterization of a black hole shadow is  discussed in \cite{Abdujabbarov:2015xqa}. The study of shadows has also been extended for the five-dimensional rotating Myers-Perry black hole \cite{Papnoi:2014aaa}. Recently, significant attention has been devoted to studying various {nonsingular} models \cite{Bambi:2014nta,Stuchlik:2014qja,Ghosh:2014mea,Schee:2015nua,Amir:2015pja,Ghosh:2015pra} including its shadow \cite{Li:2013jra,Tinchev:2015apf,Abdujabbarov:2016hnw}. 

In the absence of a well-defined quantum gravity, models of nonsingular black holes have received significant attention \cite{Ansoldi:2008jw,Bambi:2013ufa,Toshmatov:2014nya,Ghosh:2014hea}. In contrast to classical black holes where singularities are unavoidable, the nonsingular black holes {also} have horizons, but there is no singularity and their metrics as well as their curvature invariants are well behaved everywhere. Bardeen \cite{Bardeen} proposed the first nonsingular black hole which is a solution of the Einstein equations coupled to an electromagnetic field, yielding an alteration of the Reissner-Nordstr{\"o}m metric. However, the physical source associated with a Bardeen solution was obtained much later by Ay{\'o}n-Beato and Garc{\'i}a \cite{ABG99}. Later, the exact self-consistent solutions for the nonsingular black hole for the dynamics of gravity coupled to nonlinear electrodynamics was obtained  \cite{AGB,AGB1,AGB2,regular,regular1,regular2,regular3,Hayward,Culetu:2014lca,lbev,Balart:2014cga,Xiang}. Subsequently, there has been an intense activity in the investigation of nonsingular black holes \cite{regular,regular1,regular2,regular3,Hayward,Culetu:2014lca,lbev,Balart:2014cga,Xiang}. Recently, Ghosh \cite{Ghosh:2014pba} employed a probability distribution inspired mass function $m(r)$ to replace the Kerr black hole mass $ M $ to obtain a rotating nonsingular black hole. It has an additional deviation parameter $k$ which is identified asymptotically ($r>>k$) exactly as the Kerr-Newman black hole and as the Kerr black hole when $k=0$. Thus, the rotating nonsingular black hole is a generalization of the Kerr black hole. The subject of this paper is to study the size and apparent shape of this rotating nonsingular black hole, and its changes due to the parameter $k$, by analyzing in detail the unstable circular orbits. Thus, we have studied the effect of the parameter $k$ on the shape of a shadow of the Kerr black hole. As in case of the Kerr black hole, we use two observables\(-\)the radius $R_s$ and the distortion parameter $\delta_s$, characterizing the apparent shape. We found that these observables and, hence, the shape of the shadows are affected by the parameter $k$. Interestingly, the size of the shadow, decreases with parameter $k$ for a given value of $a$ in rotating nonsingular black hole, thereby suggesting a smaller shadow than in the  Kerr black hole. Thus, the larger value of the deviation parameter $k$ leads to a decrease in the size of the shadow, and thus $R_s$ decreases with an increase in $k$. The distortion of the shadow is characterized by the observable $\delta_s$ which for the case of a rotating nonsingular black hole increases monotonically with the parameter $k$. Thus, the black hole becomes more distorted or deformed with the increase in parameter $k$ for a given value of $a$.

The paper is organized as follows: In Sec.~\ref{spacetime}, we briefly review the rotating nonsingular black hole and the study of the ergoregion for this black hole. In Sec.~\ref{geodesics}, we obtain the geodesics of the photon and also discuss the effective potential. In Sec.~\ref{shadow}, we obtain the apparent shapes of the rotating nonsingular black hole and calculate the observables, and in Sec.~\ref{EE rate}, we study the energy emission rate of the black hole. Finally, we conclude in Sec.~\ref{conclusion}.

\section{Rotating nonsingular black holes}
\label{spacetime}
The metric of the rotating nonsingular black holes was obtained in Ref.~\cite{Ghosh:2014pba}, which in Boyer-Lindquist coordinates ($t$, $r$, $\theta$, $\phi$), with gravitational units $G=c=1$, reads
\begin{eqnarray}\label{metric}
ds^2 &=& - \left( 1- \frac{2Mr e^{-k/r}}{\Sigma} \right) dt^2 + \frac{\Sigma}{\Delta}dr^2 + \Sigma d \theta^2 \nonumber\\
&-& \frac{4aMr e^{-k/r}}{\Sigma  } \sin^2 \theta dt d\phi \nonumber \\
& + & \left[r^2+ a^2 + \frac{2M r a^2 e^{-k/r}}{\Sigma} \sin^2 \theta \right] \sin^2 \theta d\phi^2,
\end{eqnarray}
where $\Sigma$ and $\Delta$ are given by
\begin{eqnarray}
\Sigma = r^2 + a^2 \cos^2\theta \;\;\; \text{and}\;\;\;  \Delta=r^2 + a^2 - 2 M r e^{-k/r}.
\end{eqnarray}
The parameters $M$, $a$, and $k$ ($k>0$), respectively, represent the mass, spin, and deviation parameter. The metric (\ref{metric}) reduces to the Kerr solution \cite{Kerr:1963ud} if the parameter $k=0$ and to the Schwarzschild solution for both $k=a=0$ \cite{schw}. The detailed analysis of the horizon structure and the energy conditions of the metric (\ref{metric}) are also discussed \cite{Ghosh:2014pba}. It is noted that the metric (\ref{metric}) is singular at $\Delta=0$ and $\Sigma=0$. However, $\Delta=0$ is a coordinate singularity. It is seen that the metric~(\ref{metric}), for $r>>k$, can be identified as Kerr-Newman \cite{Newman:1965my}, i.e., for $r>>k$, one gets
\begin{eqnarray}
g_{tt}&=&1-\frac{2Mr-q^2}{\Sigma}+\mathcal{O}(k^2/r^2),\nonumber \\
\Delta &=& r^2+a^2-2Mr+q^2+\mathcal{O}(k^2/r^2), \nonumber
\end{eqnarray}
where the charge $q$ and mass $M$ are related to the deviation parameter $k$ via $q^2=2Mk$. 

The astrophysical black hole  candidates are expected to be the Kerr black holes, but the actual nature of these astrophysical objects has still to be verified \cite{Bambi:2011mj}. The nonsingular metric can be seen as the prototype of a large class of non-Kerr black hole metrics, in which the components of the metric tensor  have the same expressions as that of the Kerr black hole with $M$ replaced by a mass function $ m(r) $ that reduces to $ M $ as $r \rightarrow 0$ \cite{Bambi:2013sha}. Further, non-Kerr black holes or nonsingular ones may indeed look like Kerr black holes with different spin; e.g., the nonsingular ones with some nonzero values of the deviation parameter and spin parameter that are higher (lower) than the value of the spin parameter (say $a^*$) look like Kerr black holes with spin parameter higher (lower) than $a^*$. These black holes have  an additional deviation parameter. The standard procedure to test the Kerr black hole with the observational x-ray data in Cygnus X-1 is developed in Refs. \cite{Gou:2011nq,Gou:2013dna}.  Thus, to test the nonsingular metric with the x-ray data of the black hole candidate in Cygnus X-1 would be to repeat the algorithm of Refs. \cite{Gou:2011nq,Gou:2013dna} with the Kerr background replaced by the nonsingular black holes. 

The procedure has been applied to test the Bardeen metric with the x-ray data of the black hole candidate in Cygnus X-1 with the Kerr black hole background replaced by the Bardeen \cite{Bambi:2014nta}. The spin parameter and the inclination angle can be determined by observing the apparent shape of the shadow, which is distorted mainly by these two parameters, defining some observables characterizing the apparent shape (its radius and distortion parameter). It has been shown that the spin parameter and the inclination angle of a Kerr black hole can be determined by the observation \cite{Hioki:2009na}. One can assume that the observations may allow non-Kerr black holes or the nonsingular black holes considered as real astrophysical black hole candidates; hence, the technique \cite{Hioki:2009na} can be used to determine how the spin parameter may change due to the additional parameter $k$.

\begin{figure*}
	\begin{tabular}{c c c c}
 \includegraphics[width=0.245\linewidth]{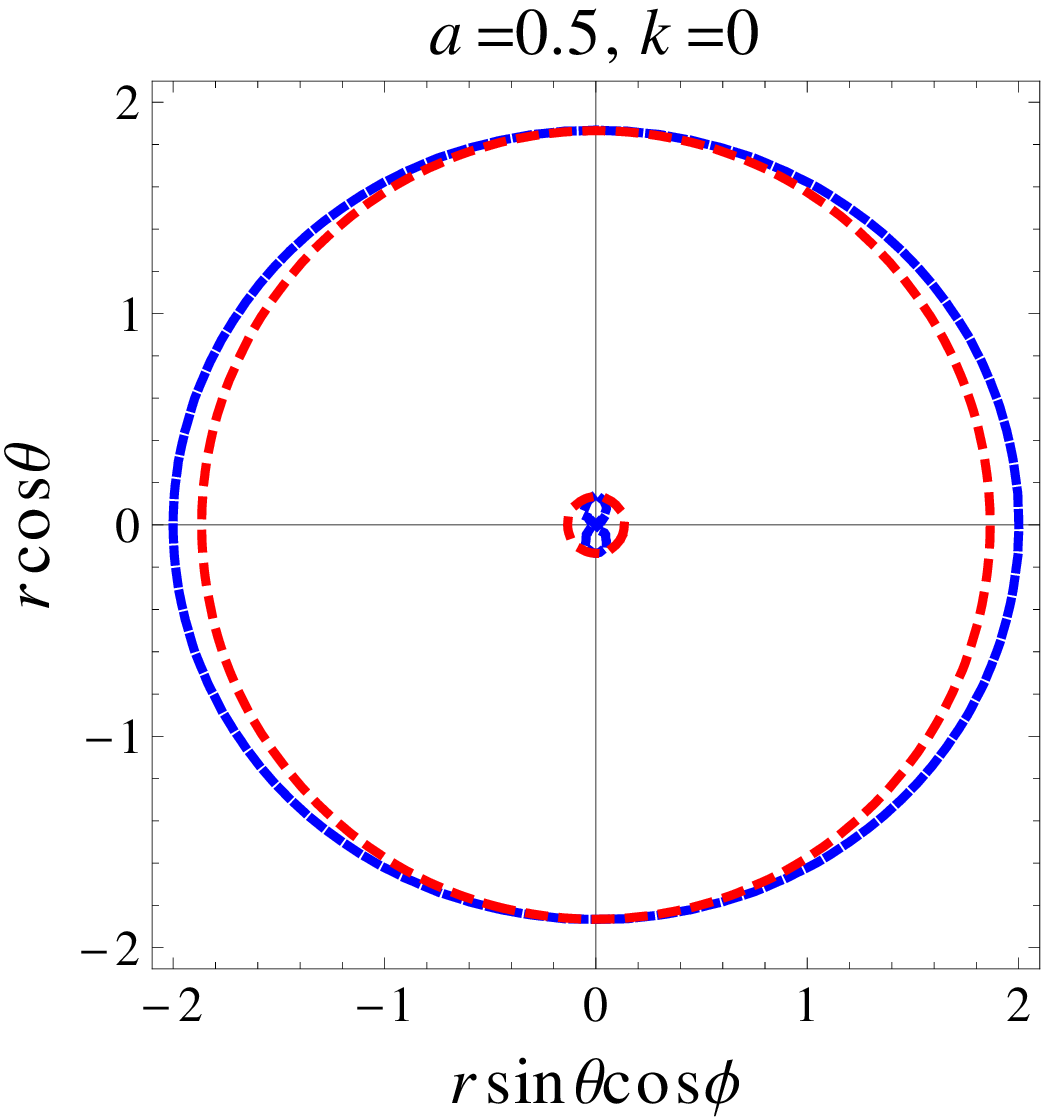}
 \includegraphics[width=0.245\linewidth]{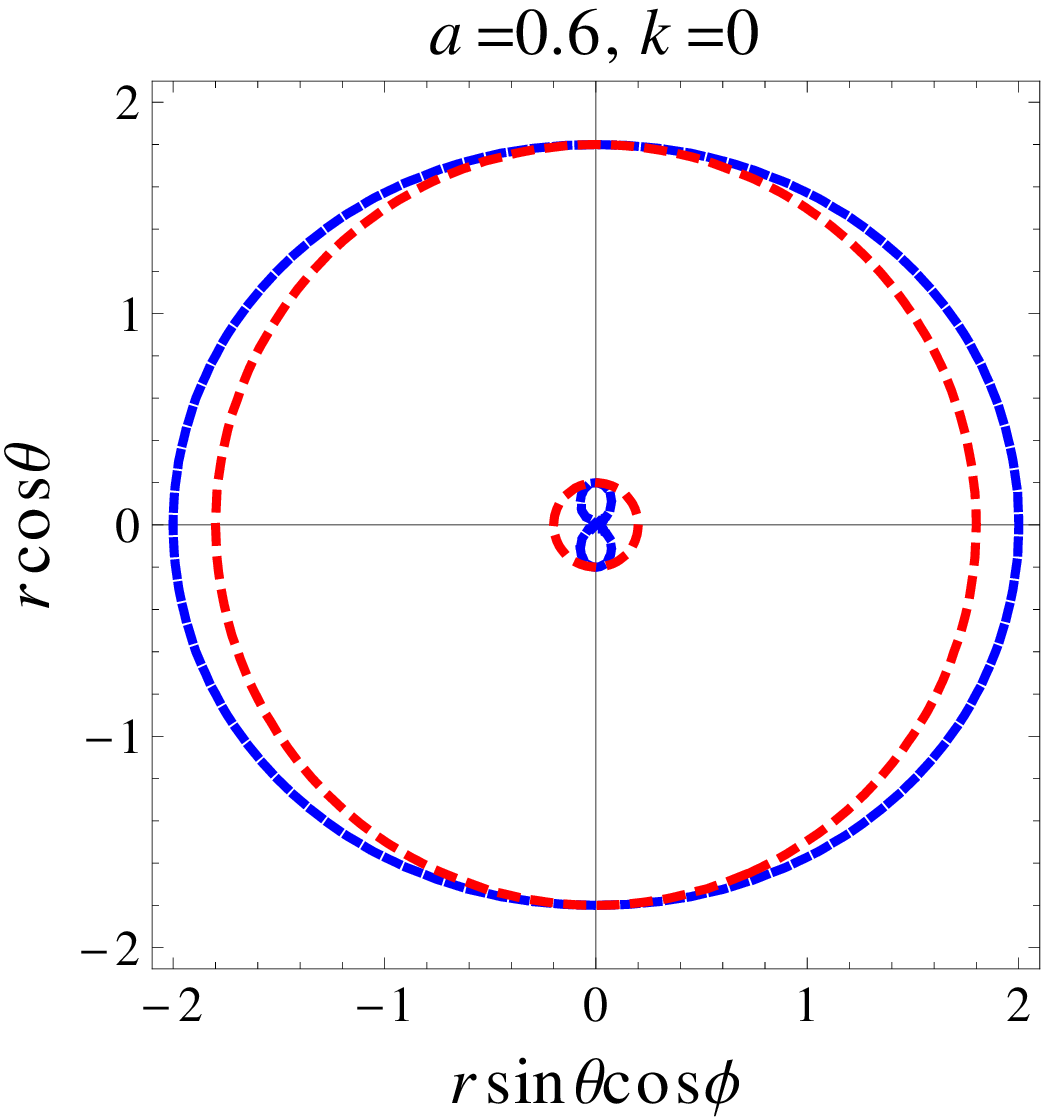}
 \includegraphics[width=0.245\linewidth]{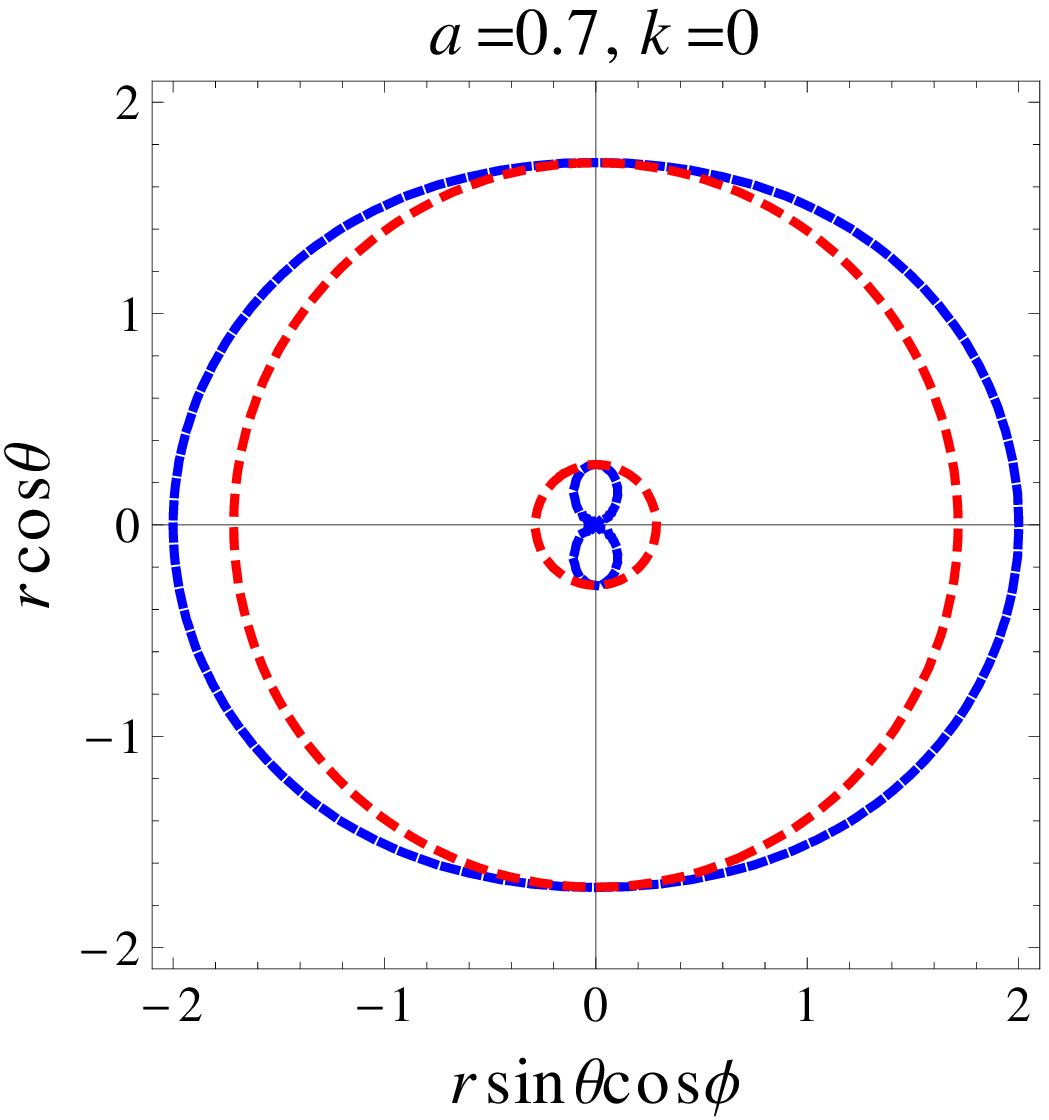}
 \includegraphics[width=0.245\linewidth]{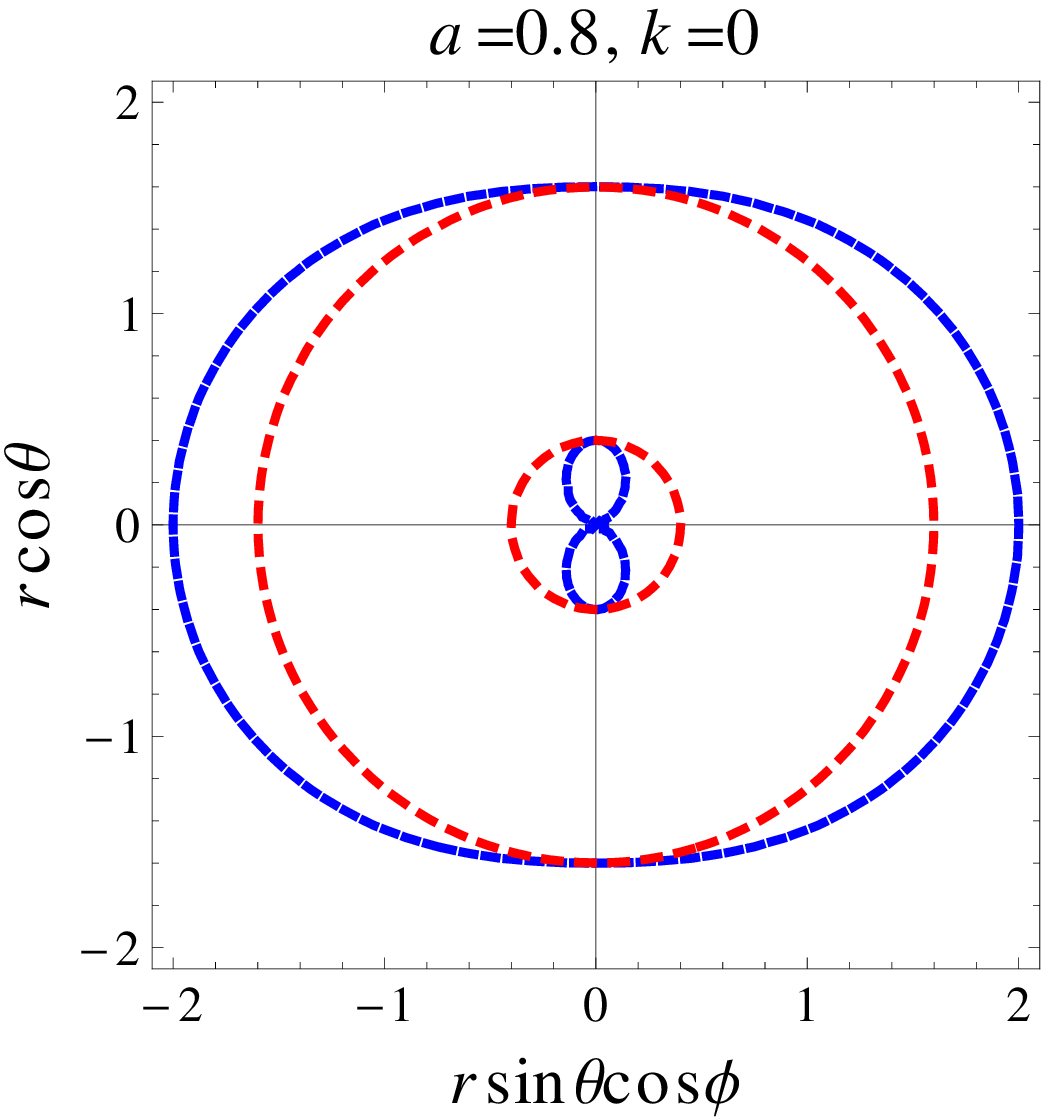}
	\end{tabular}
\caption{\label{ergo2} Plot showing the ergoregions in x-z plane with parameter $k=0$ (Kerr black hole) for various values of $a$. The blue and red lines indicating the static limit surface and horizons, respectively. The outer blue line indicates static the limit surface, while the two red lines correspond to the event and Cauchy horizons.}
\end{figure*}
\begin{figure*}
	\begin{tabular}{c c c c}
 \includegraphics[width=0.245\linewidth]{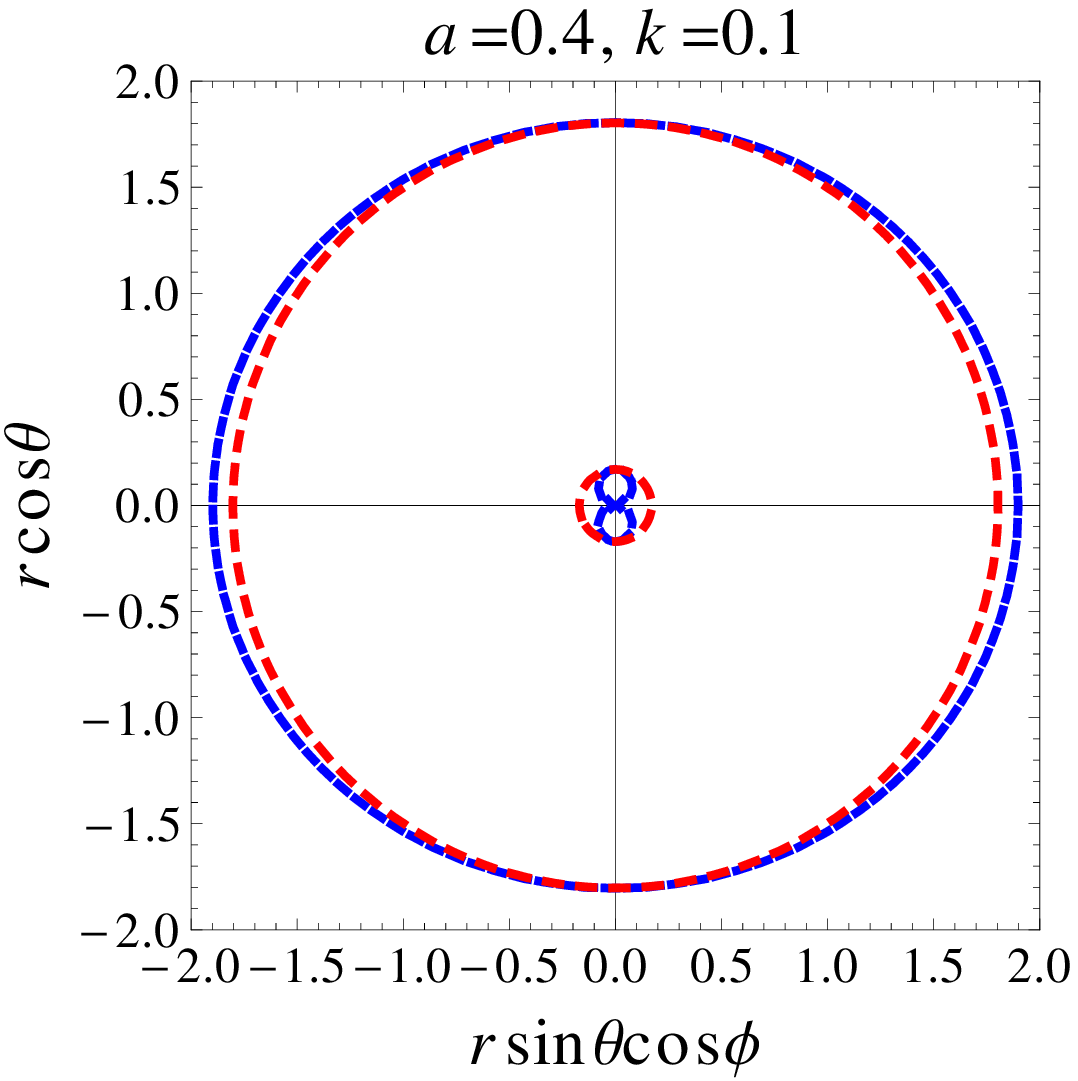}
 \includegraphics[width=0.245\linewidth]{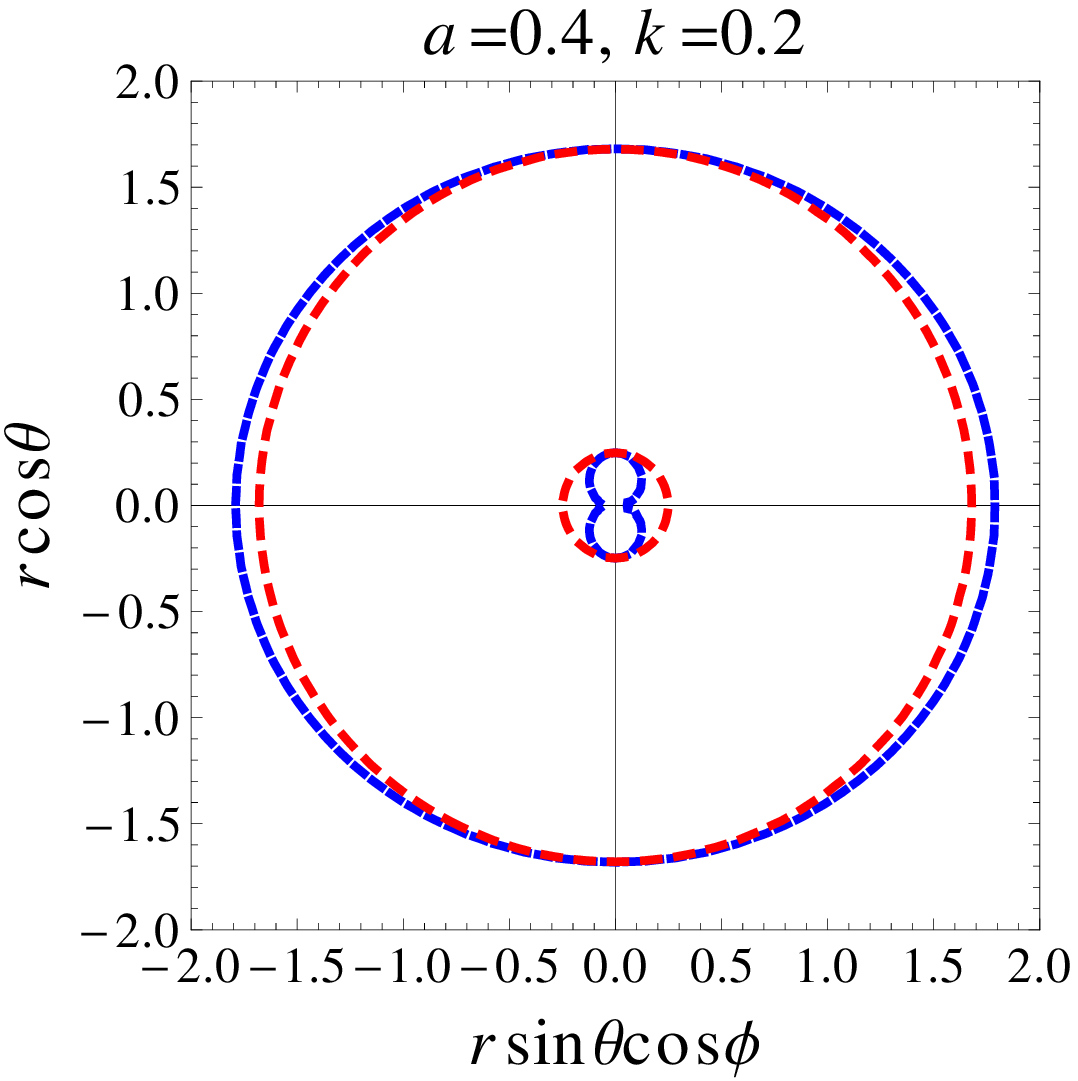}
 \includegraphics[width=0.245\linewidth]{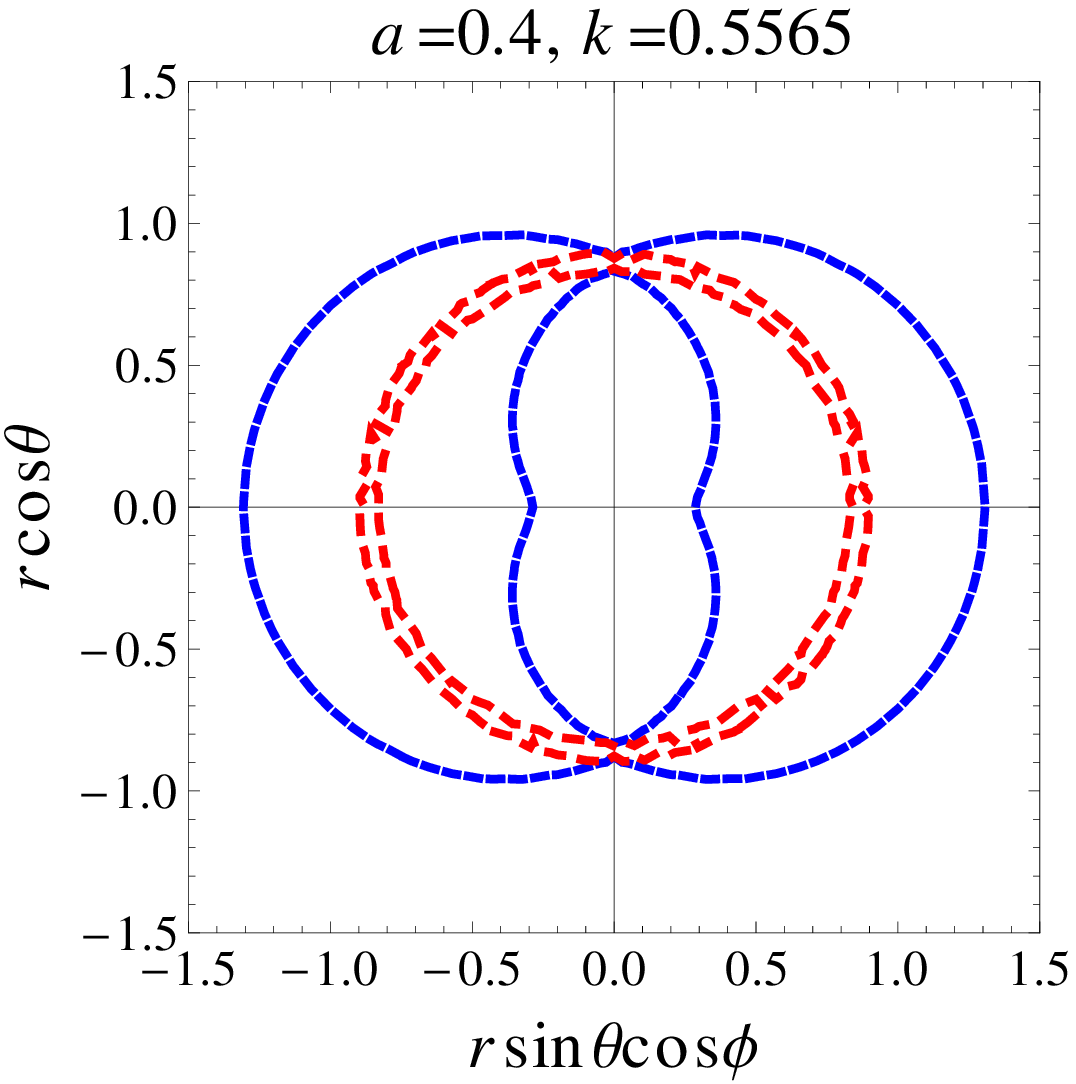}
 \includegraphics[width=0.245\linewidth]{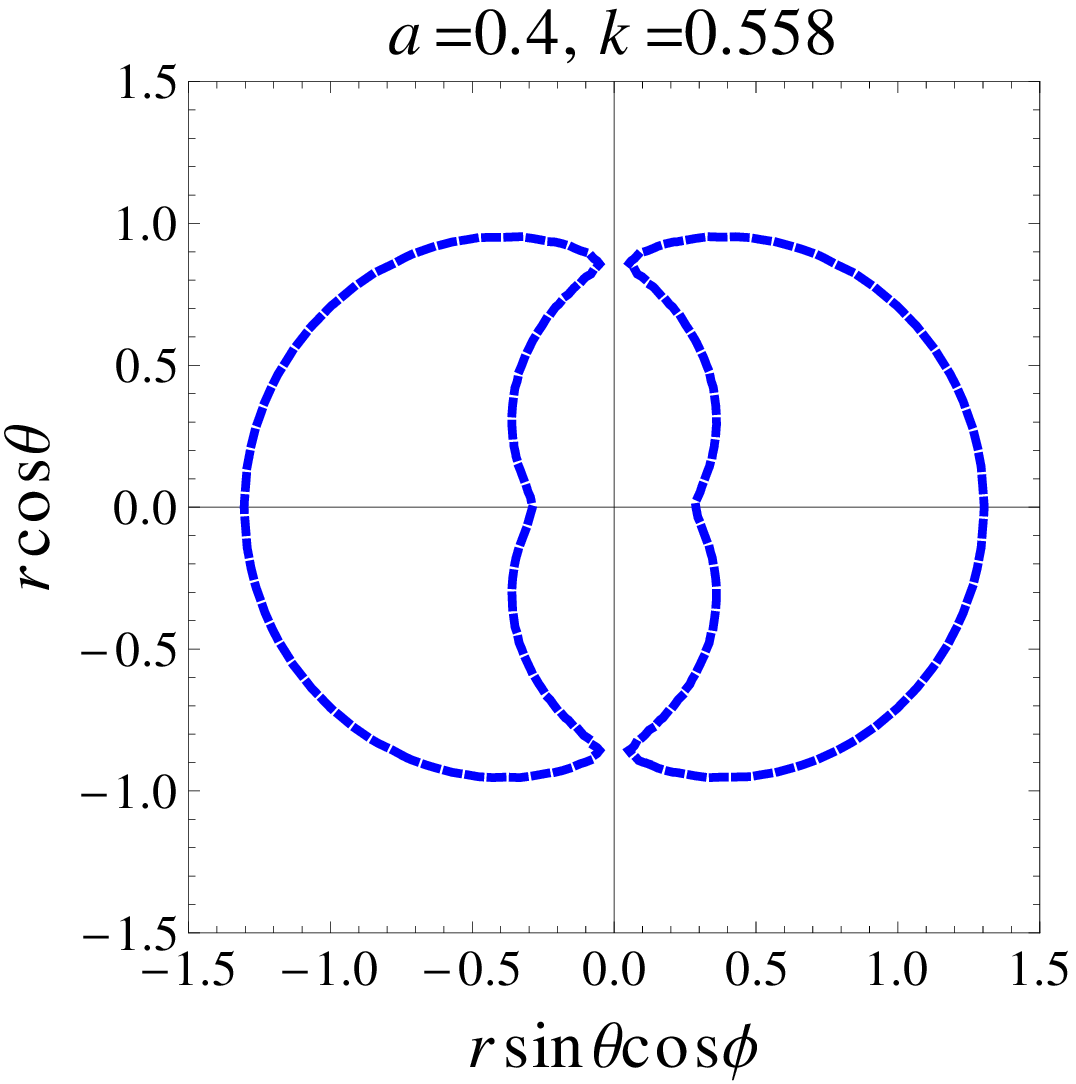}\\
 \includegraphics[width=0.245\linewidth]{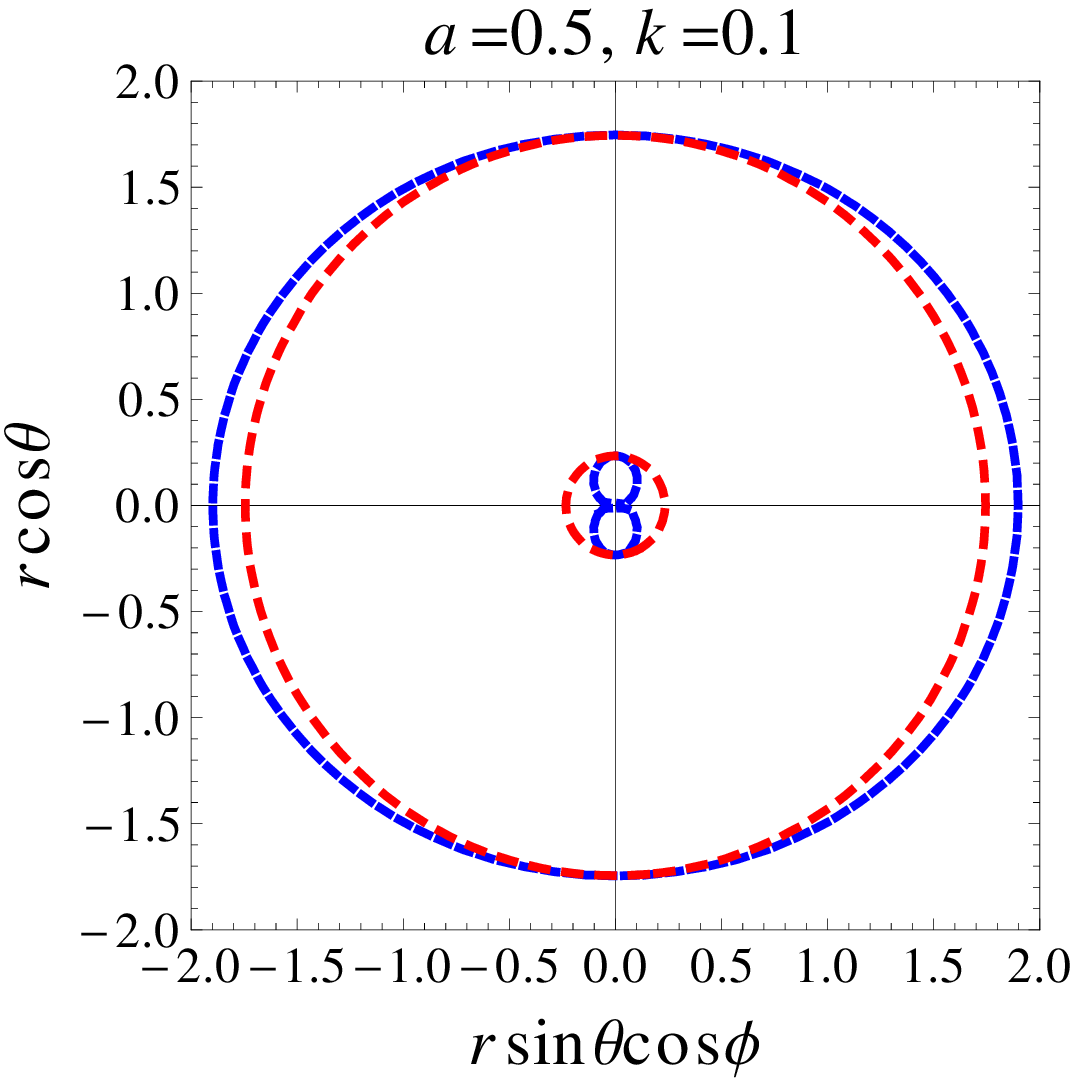}
 \includegraphics[width=0.245\linewidth]{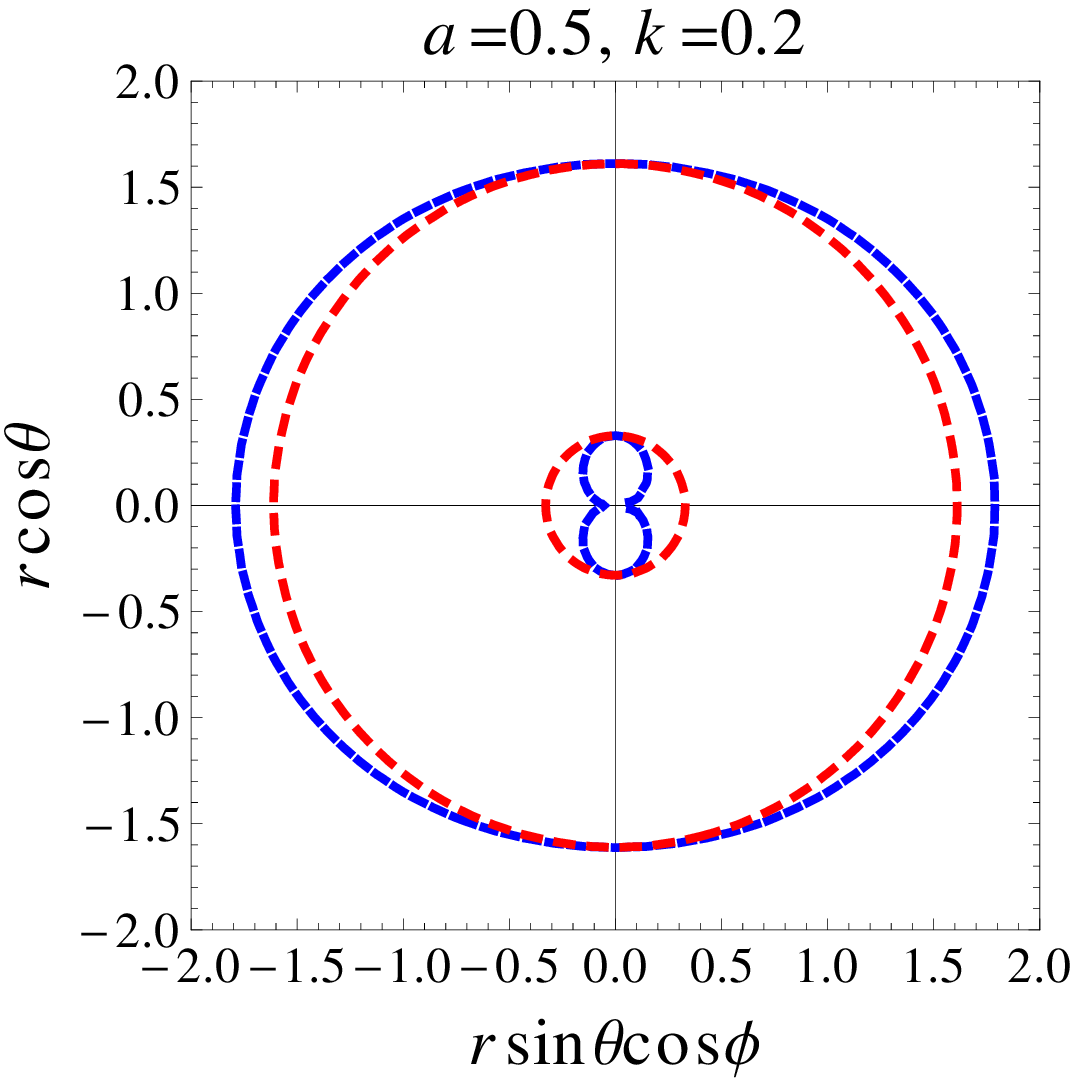}
 \includegraphics[width=0.245\linewidth]{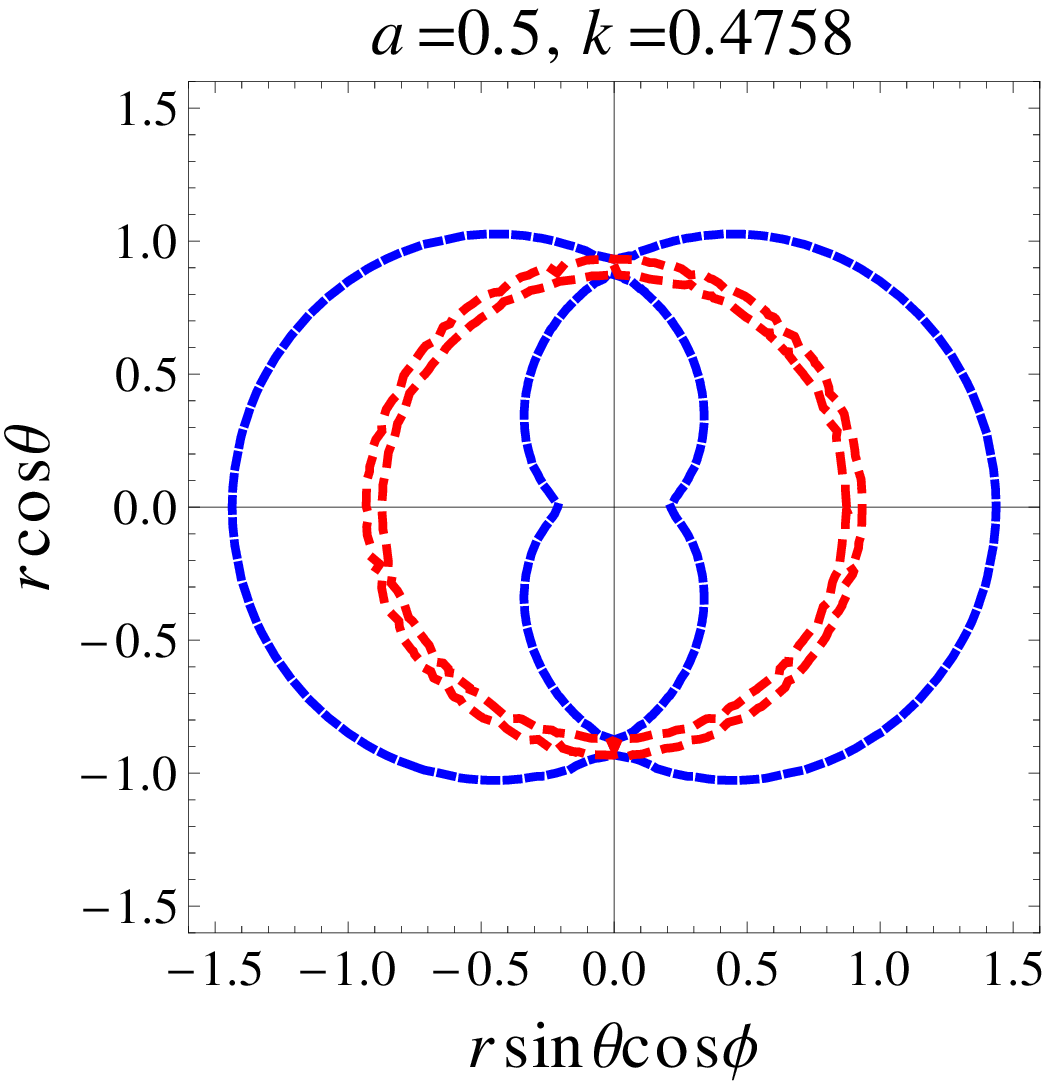}
 \includegraphics[width=0.245\linewidth]{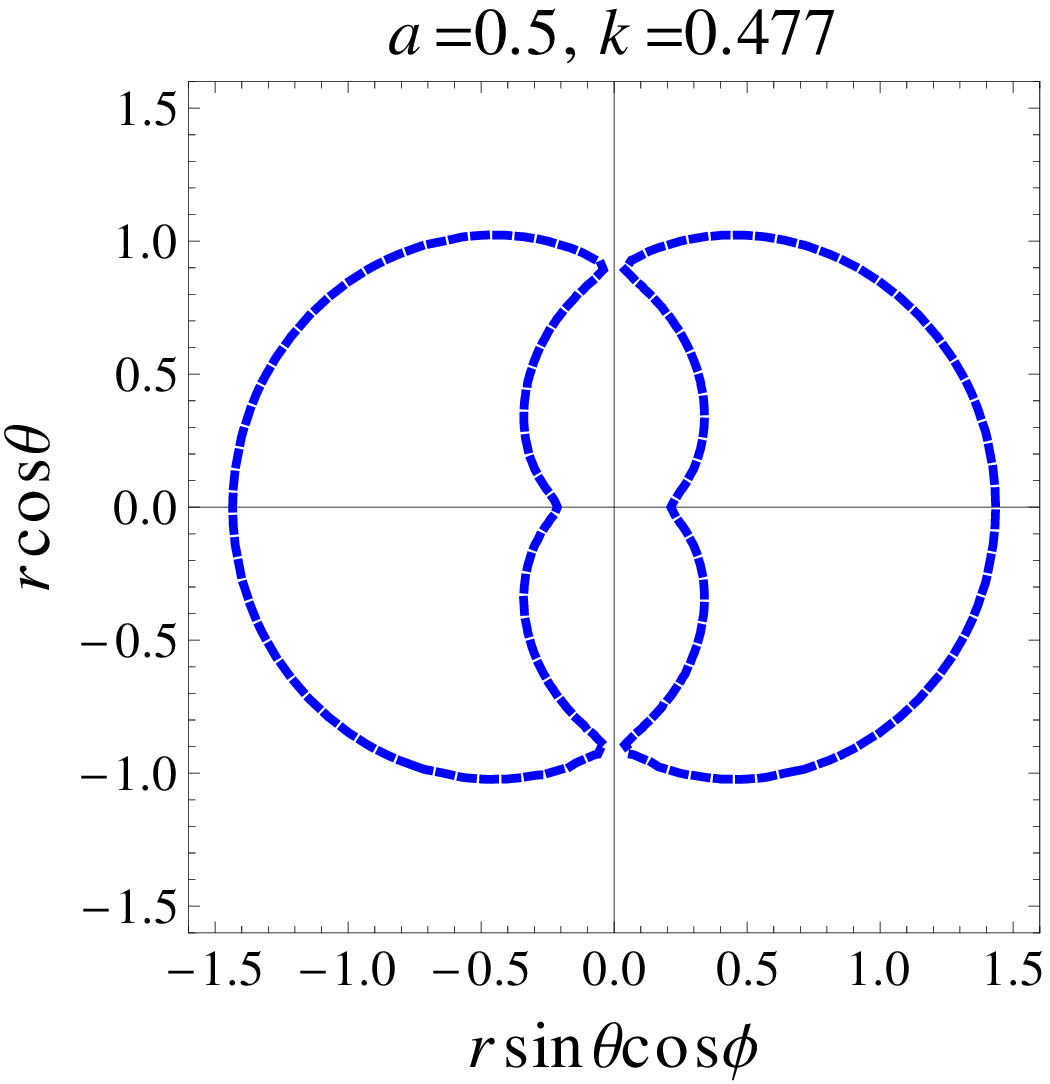}\\
 \includegraphics[width=0.245\linewidth]{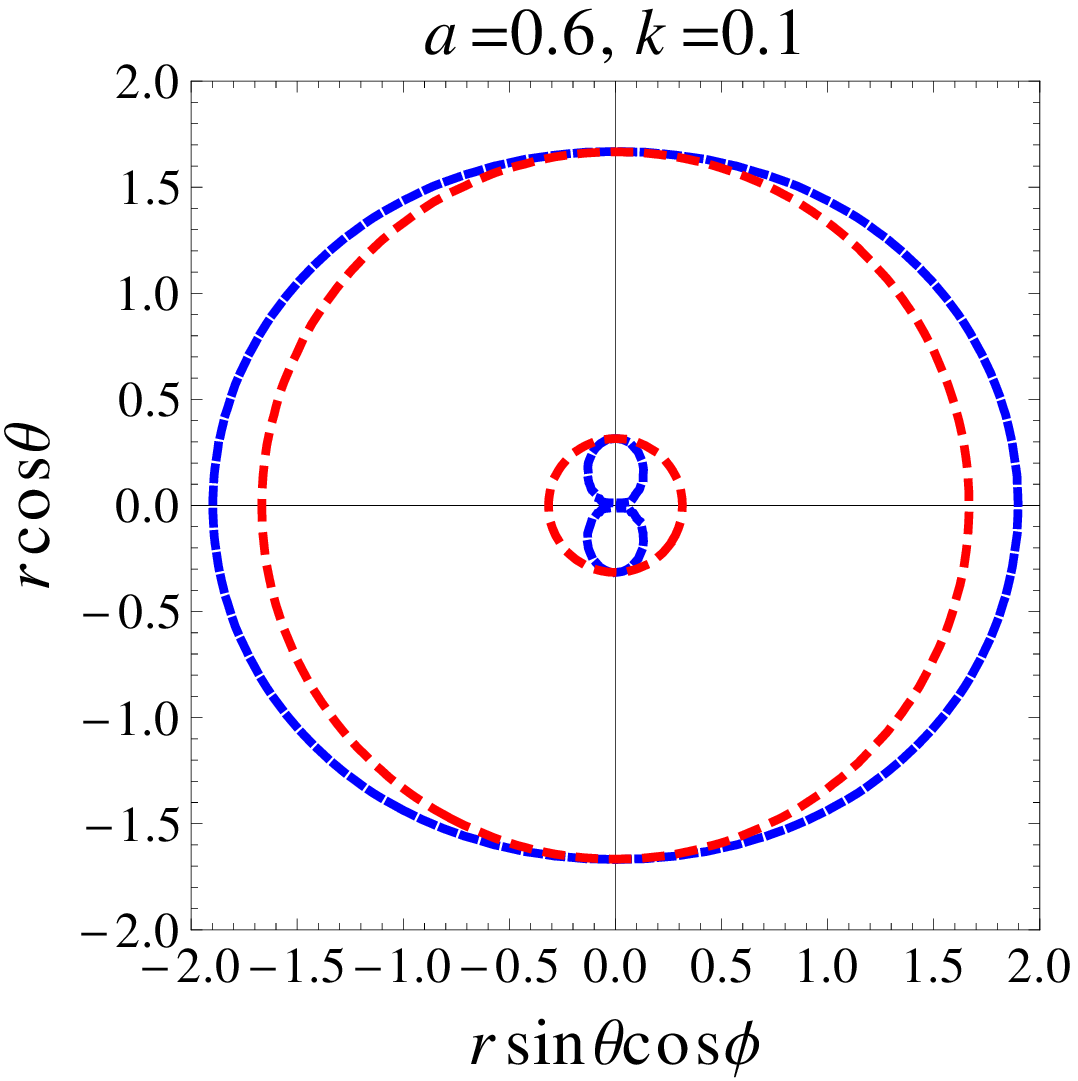}
 \includegraphics[width=0.245\linewidth]{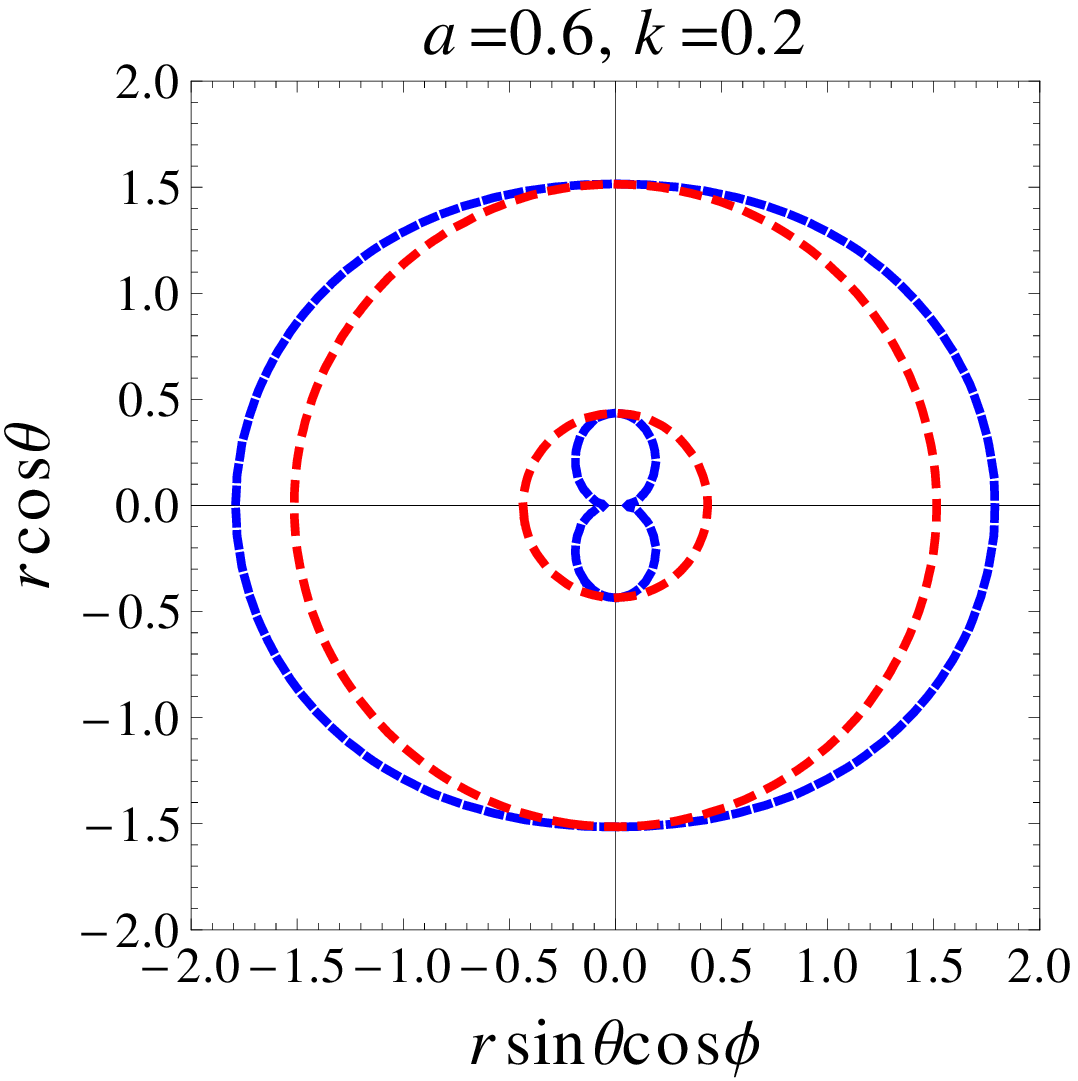}
 \includegraphics[width=0.245\linewidth]{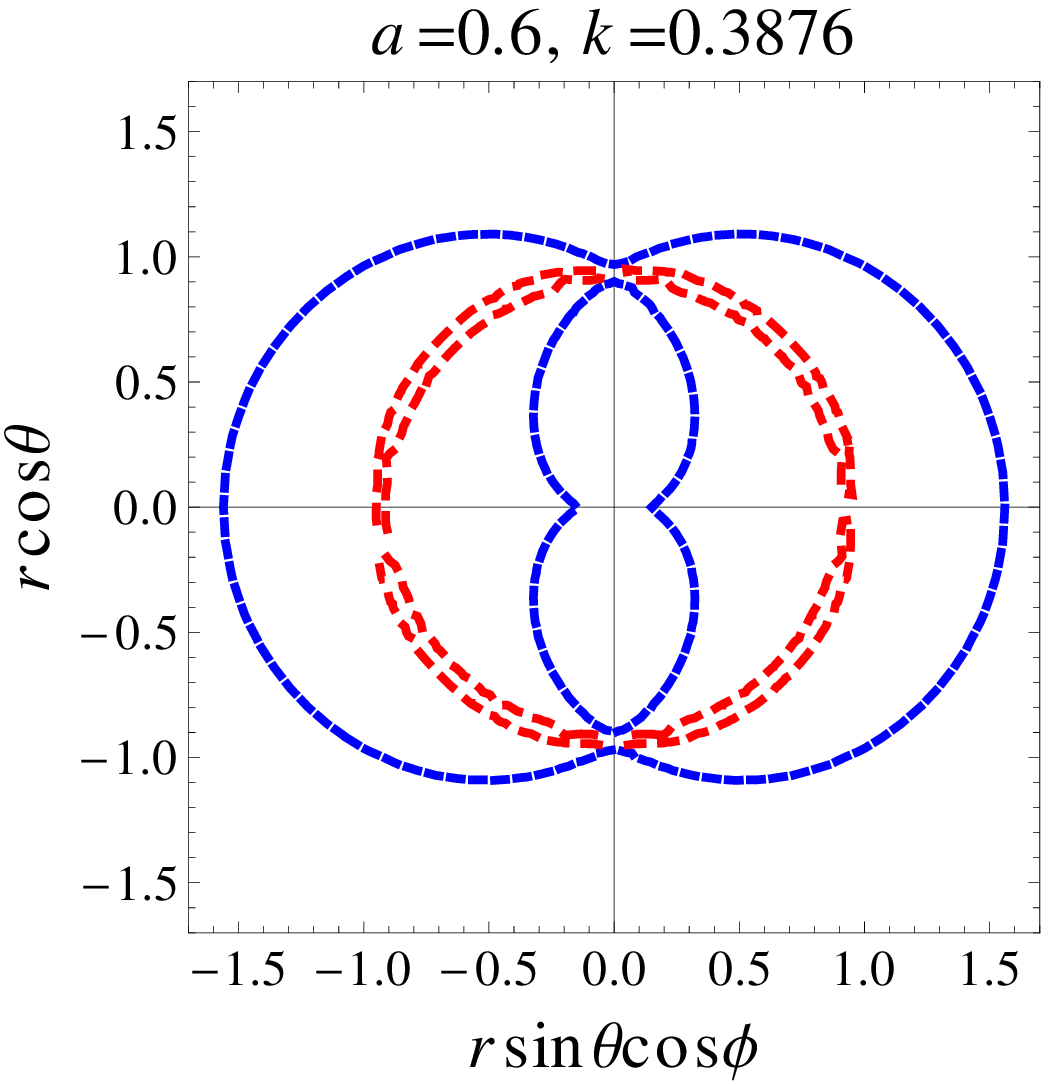}
 \includegraphics[width=0.245\linewidth]{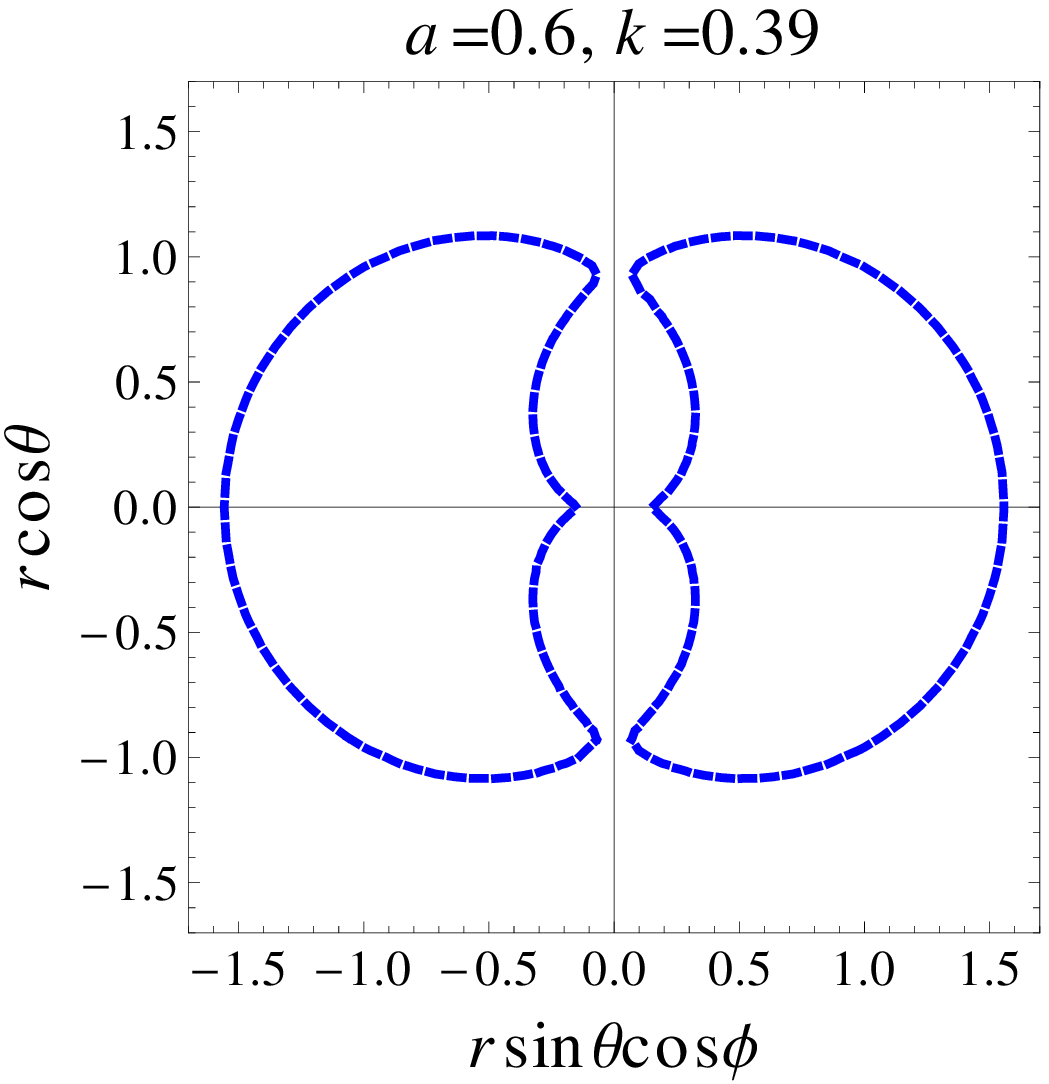}\\
 \includegraphics[width=0.245\linewidth]{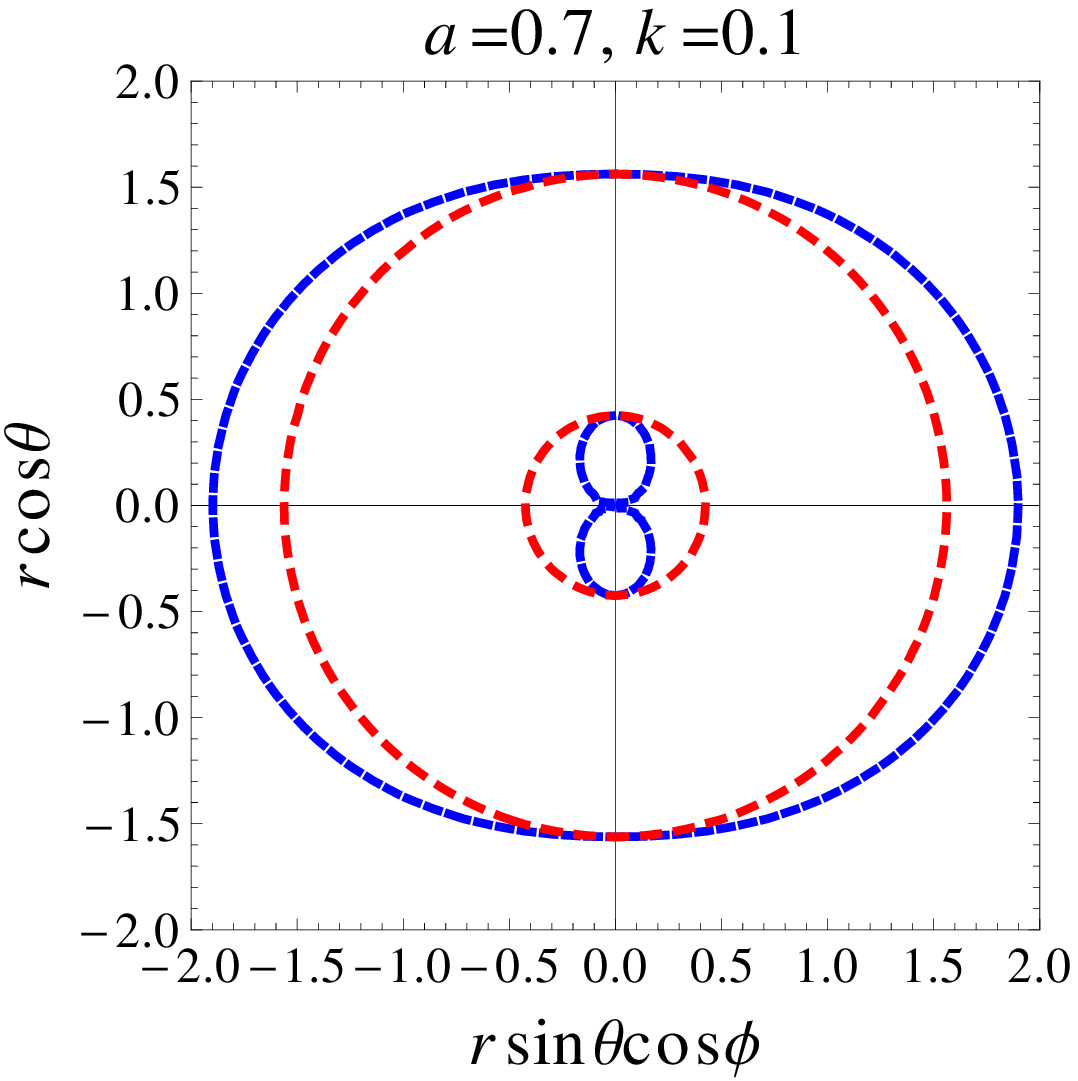}
 \includegraphics[width=0.245\linewidth]{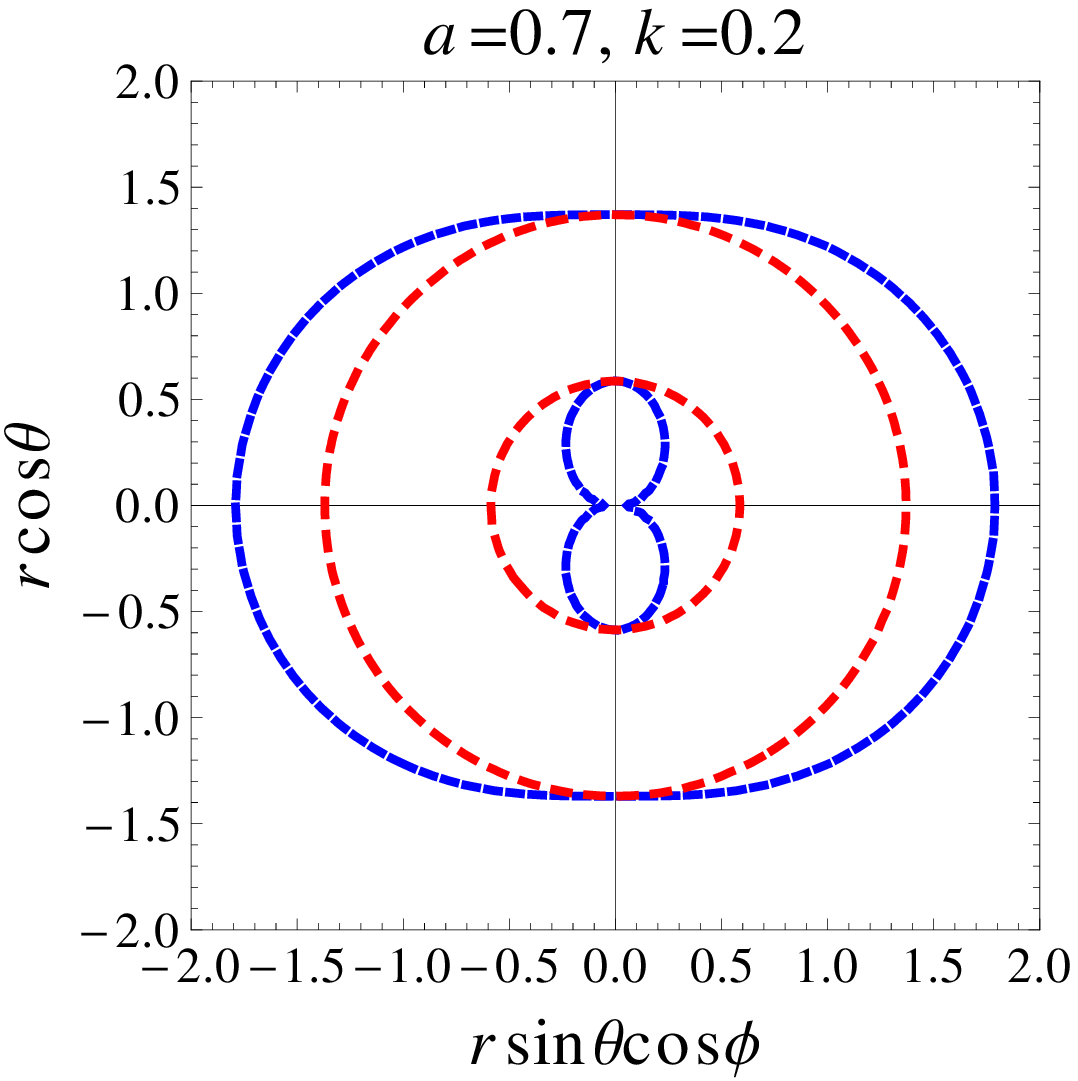}
 \includegraphics[width=0.245\linewidth]{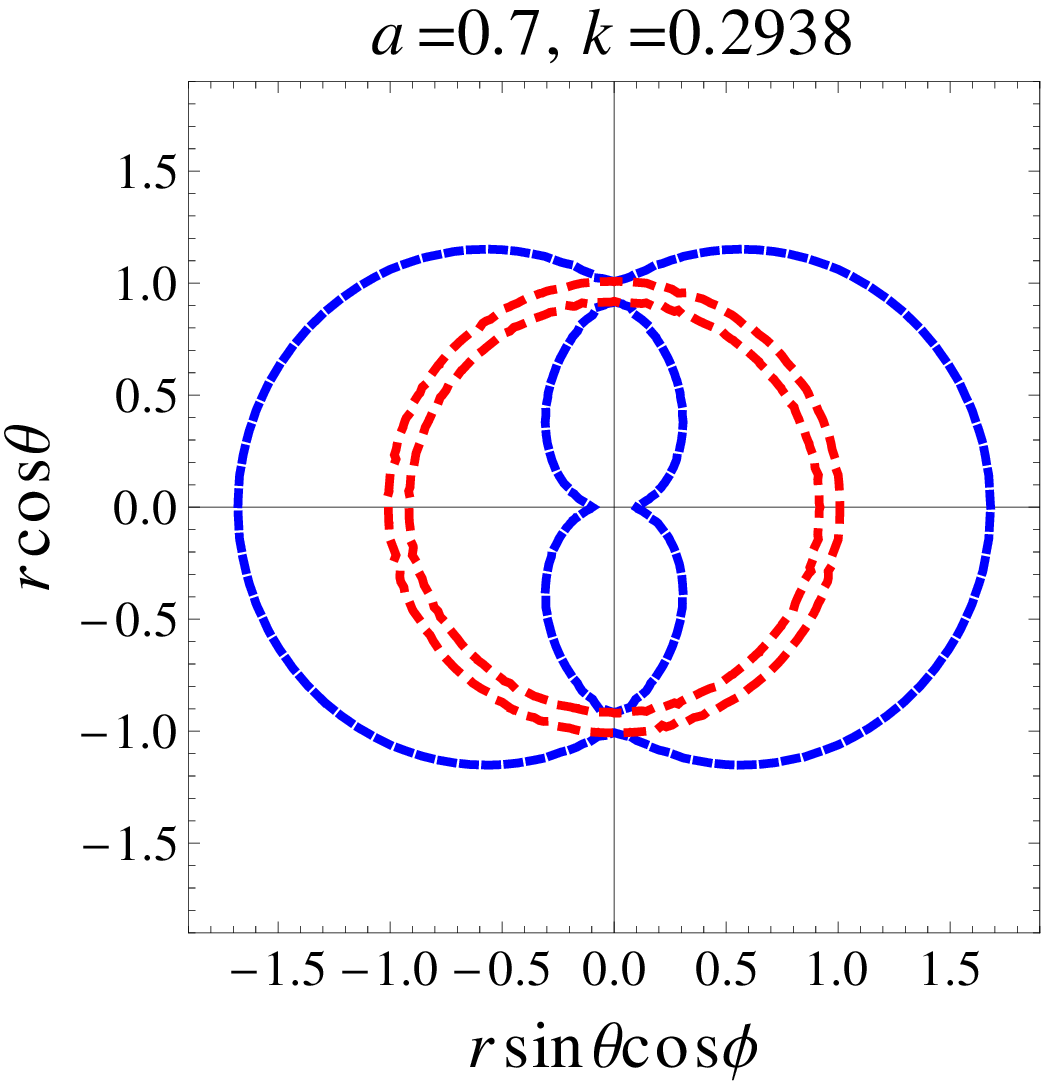}
 \includegraphics[width=0.245\linewidth]{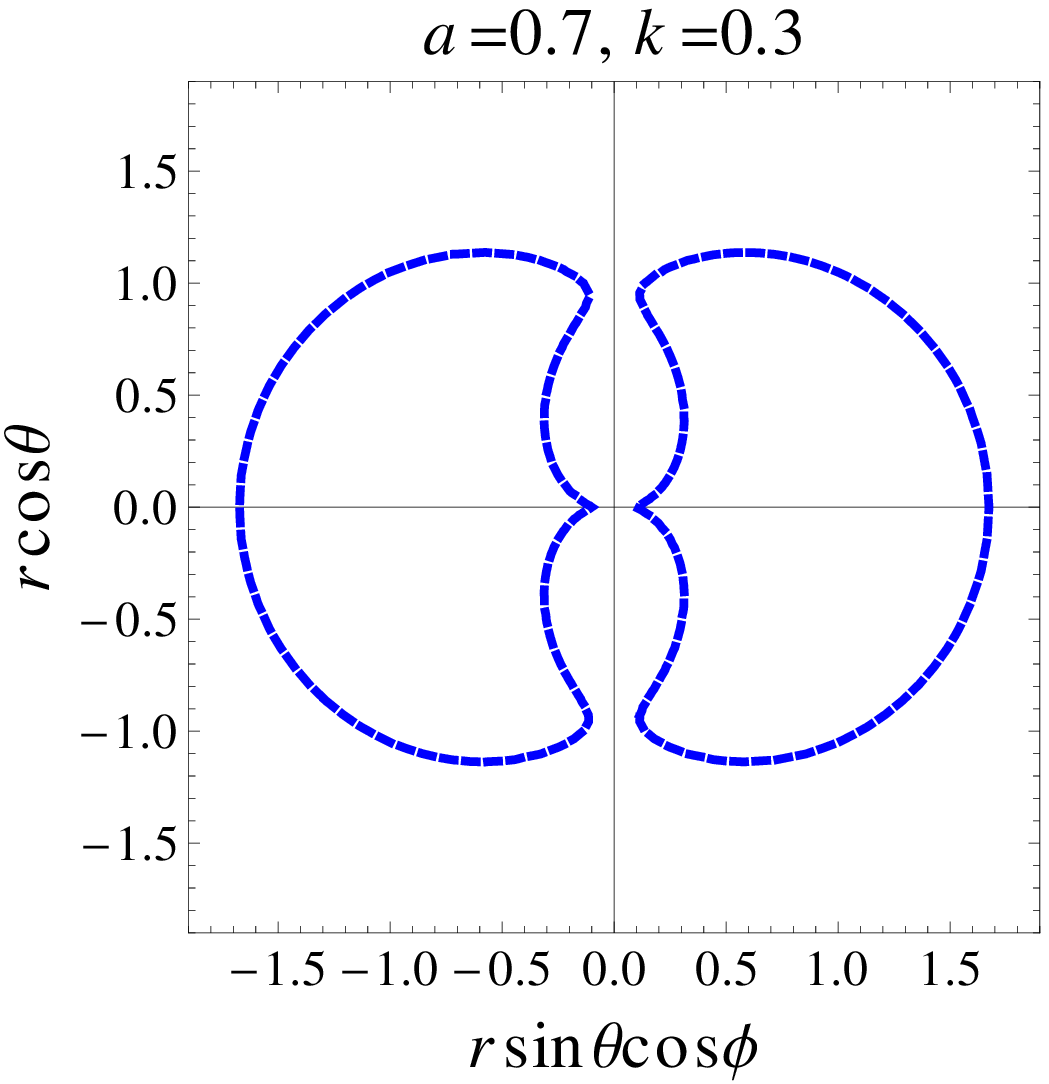}\\
 \includegraphics[width=0.245\linewidth]{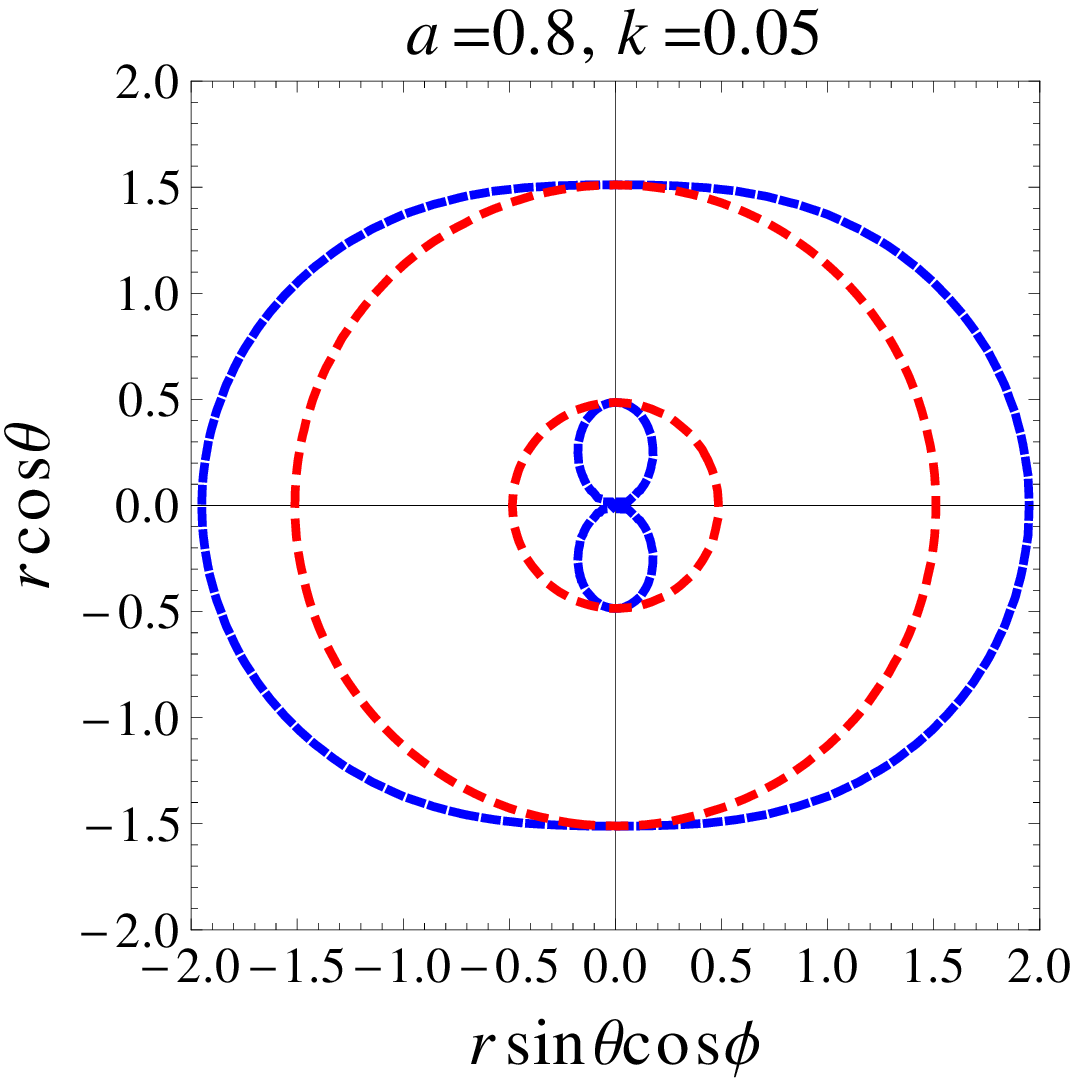}
 \includegraphics[width=0.245\linewidth]{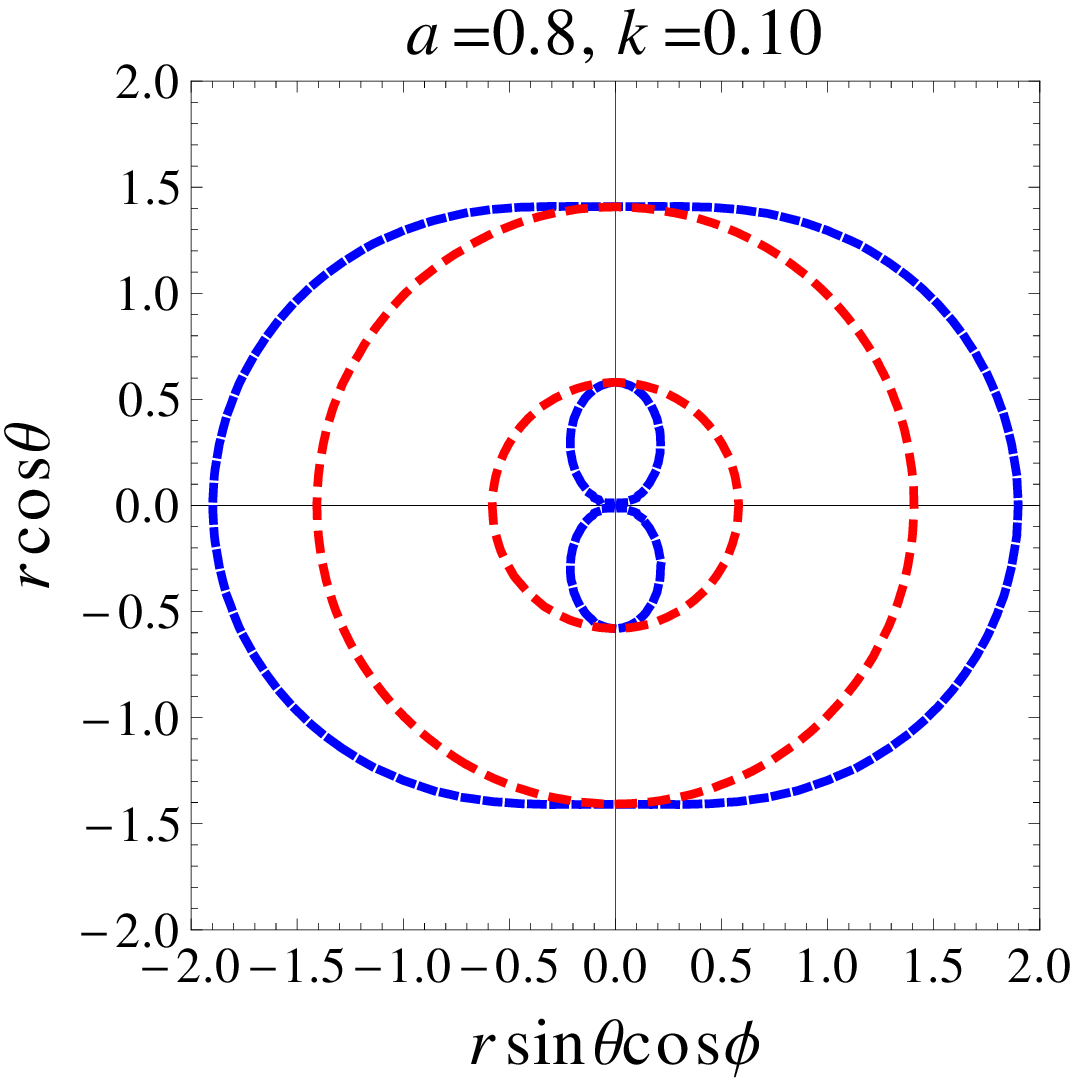}
 \includegraphics[width=0.245\linewidth]{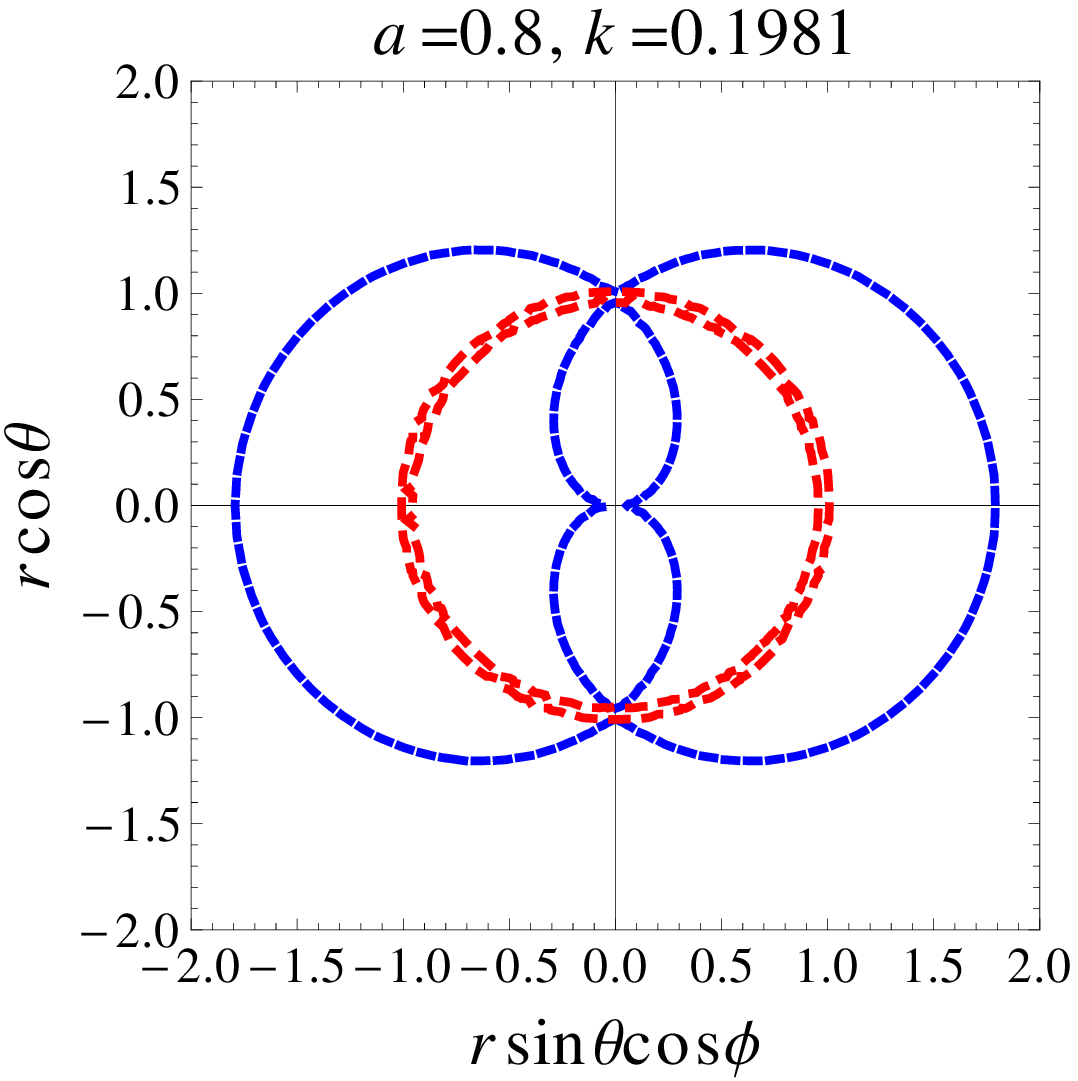}
 \includegraphics[width=0.245\linewidth]{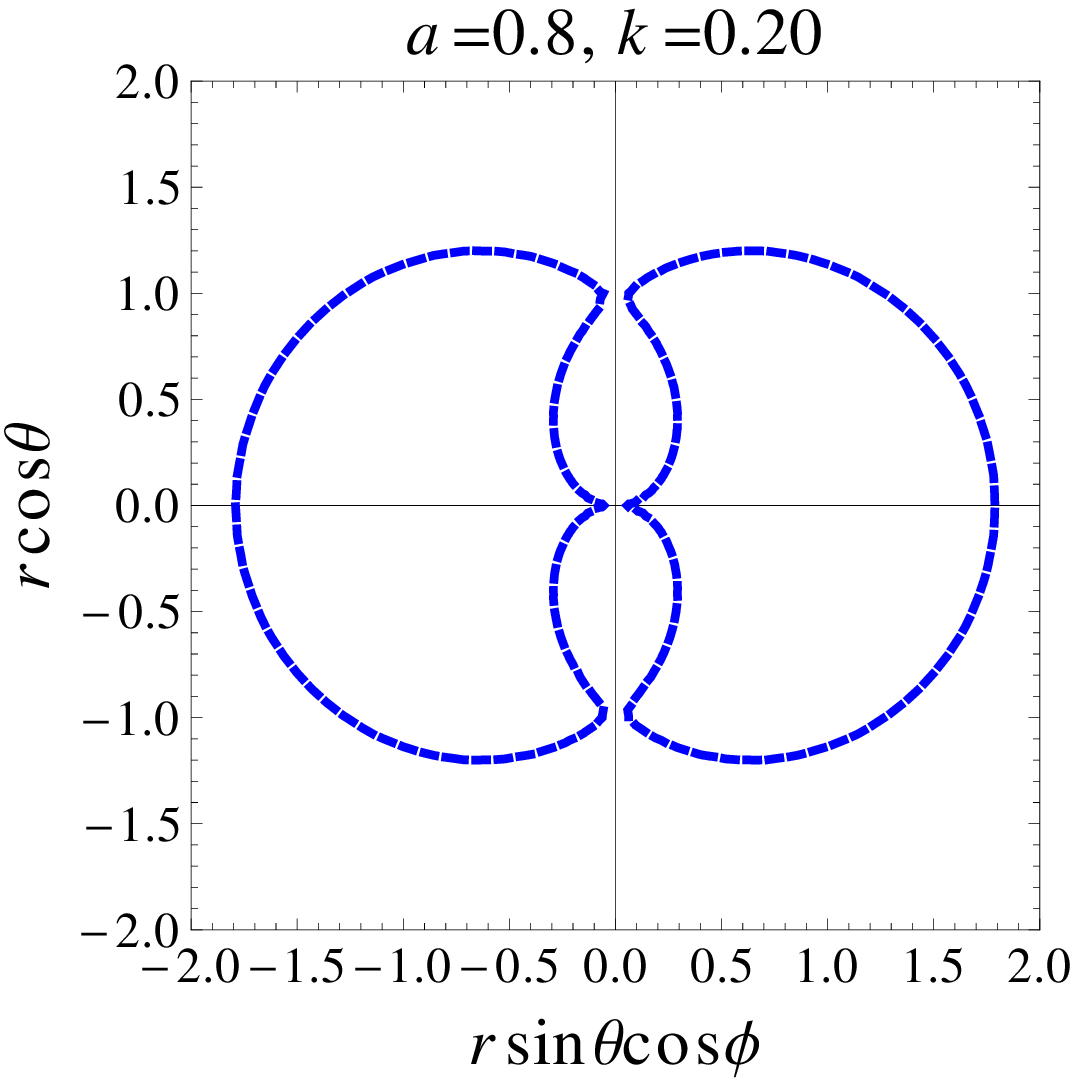}
	\end{tabular}
\caption{\label{ergo} Plot showing the ergoregions in x-z plane with parameter $a$ for various values of $k$. The blue and red lines indicating the static limit surface and horizons, respectively. The outer blue line indicates the static limit surface, while the two red lines correspond to the event and Cauchy horizons.}
\end{figure*}
\subsection{Ergoregion}
The nonsingular rotating black hole  metric (\ref{metric}) or any other rotating nonsingular black hole, e.g., the rotating Bardeen black hole \cite{Bambi:2008jg}, can be thought of as a non-Kerr black hole metric with $M$ replaced by $m(r)$, and it reduces to the standard Kerr black hole metric as $r \rightarrow \infty.$  Hence, the event horizon of the black hole is a null surface determined by $g^{rr}=\Delta=0$, and the zeros of
\begin{equation}\label{hor}
r^2 + a^2 - 2 M r e^{-k/r} =0,
\end{equation}
give the black hole horizons. The numerical solution of Eq.~(\ref{hor}) reveals that, for a suitable choice of parameters, it admits two roots, $r^{-}$ and $r^{+}$, which correspond to the Cauchy and event horizons, respectively \cite{Ghosh:2014pba}. Furthermore, it is worthwhile to understand that one can find values of parameters such that $r^{+}=r^{-}$, which means one gets an extremal black hole, and $r^{+}>r^{-}$ describes a nonextremal black hole. The horizons exist only for a certain range of the parameters \cite{Ghosh:2014pba}.
On the other hand, at the static limit surface, the timelike Killing vector $\xi^{a}=(\frac{\partial}{\partial t})^{a}$ becomes null, i.e.,
\begin{equation}
\xi^{a}\xi_{a}=g_{tt}=0,
\end{equation}
or
\begin{equation}\label{sls}
r^2 + a^2 \cos^2 \theta - 2 M r e^{-k/r} =0.
\end{equation}
By solving Eq.~(\ref{sls}) numerically, it turns out that the metric (\ref{metric}) admits two positive roots (cf. \cite{Ghosh:2014pba} for details). The region between $r^{+}_{EH} < r < r^{+}_{SLS}$ is called the ergoregion. In an ergoregion the gravitational forces change the movement of an object, and an object in this region can no longer remain stationary. In this region, an object can escape the gravitational forces of the black hole if the speed of this object is greater than the escape velocity. One of the important properties of an ergoregion is that, in this region, a timelike Killing vector $\partial_{t}$ becomes spacelike or vice versa. It is known that an ergoregion plays a significant role in the energy extraction process or Penrose process of a black hole \cite{Penrose:1971uk}.

The plots of ergoregions of the Kerr black hole ($k=0$) can be seen from Fig.~\ref{ergo2} for different values of spin $a$, showing that the ergoregion area increases with $a$. We have shown the effect of parameter $k$ on the shape of an ergoregion in Fig.~\ref{ergo}. It can be observed from these figures that the ergoregion is easily affected by the parameter $k$. It can be seen from Fig.~\ref{ergo} that one can find a critical value of $k$ ($k^c$) where the two horizons almost coincide (near extremal black hole). It can be observed for $k>k^c$ that the horizons {vanish} (cf., Fig.~\ref{ergo}). One can easily observe from Fig.~\ref{ergo} that there is an increase in the ergoregion area with the parameter $k$ for a given value of spin $a$. It turns out that the critical values of parameter $k^{c}=0.5565, 0.4758, 0.3876, 0.2938, 0.1981$ correspond, respectively, for $a=0.4, 0.5, 0.6, 0.7, 0.8$.

\section{Geodesics in nonsingular black hole}
\label{geodesics}
\begin{figure*}
	\begin{tabular}{c c}
	 \includegraphics[width=0.5\linewidth]{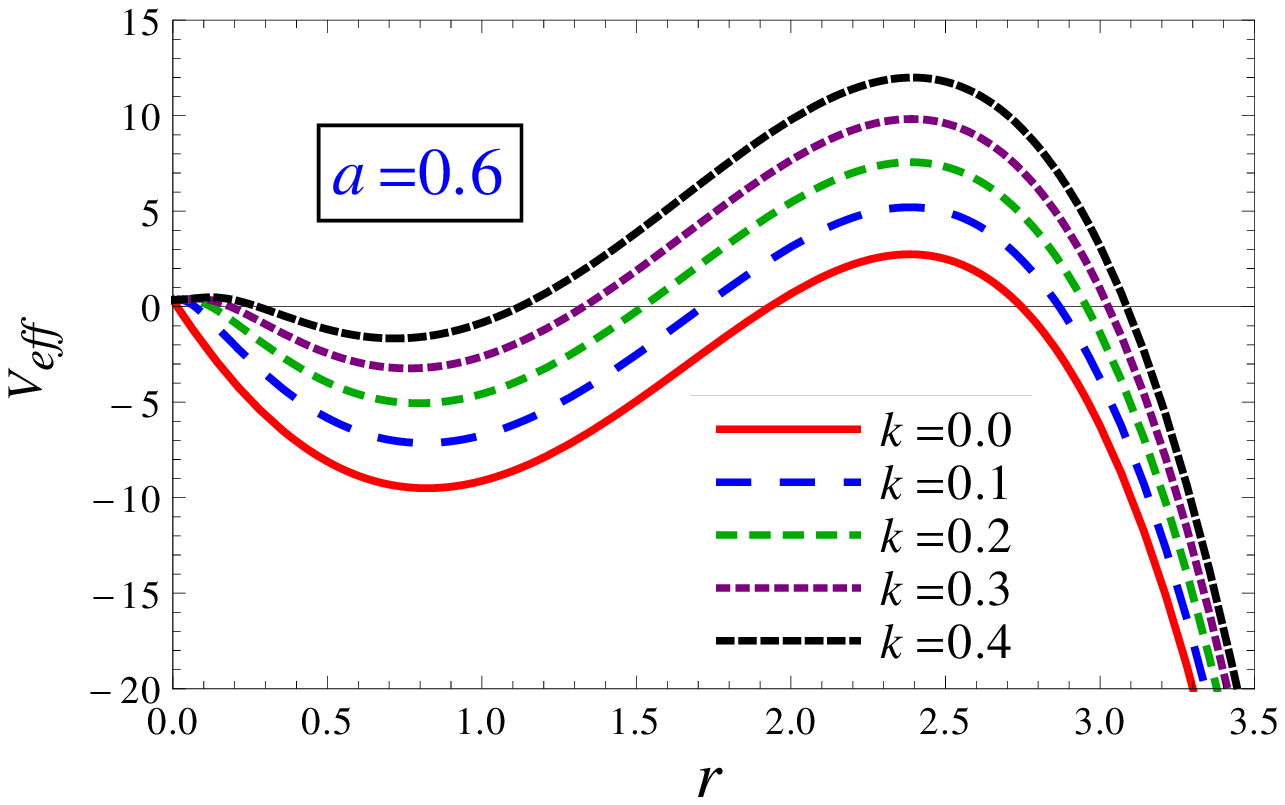}
	 \includegraphics[width=0.5\linewidth]{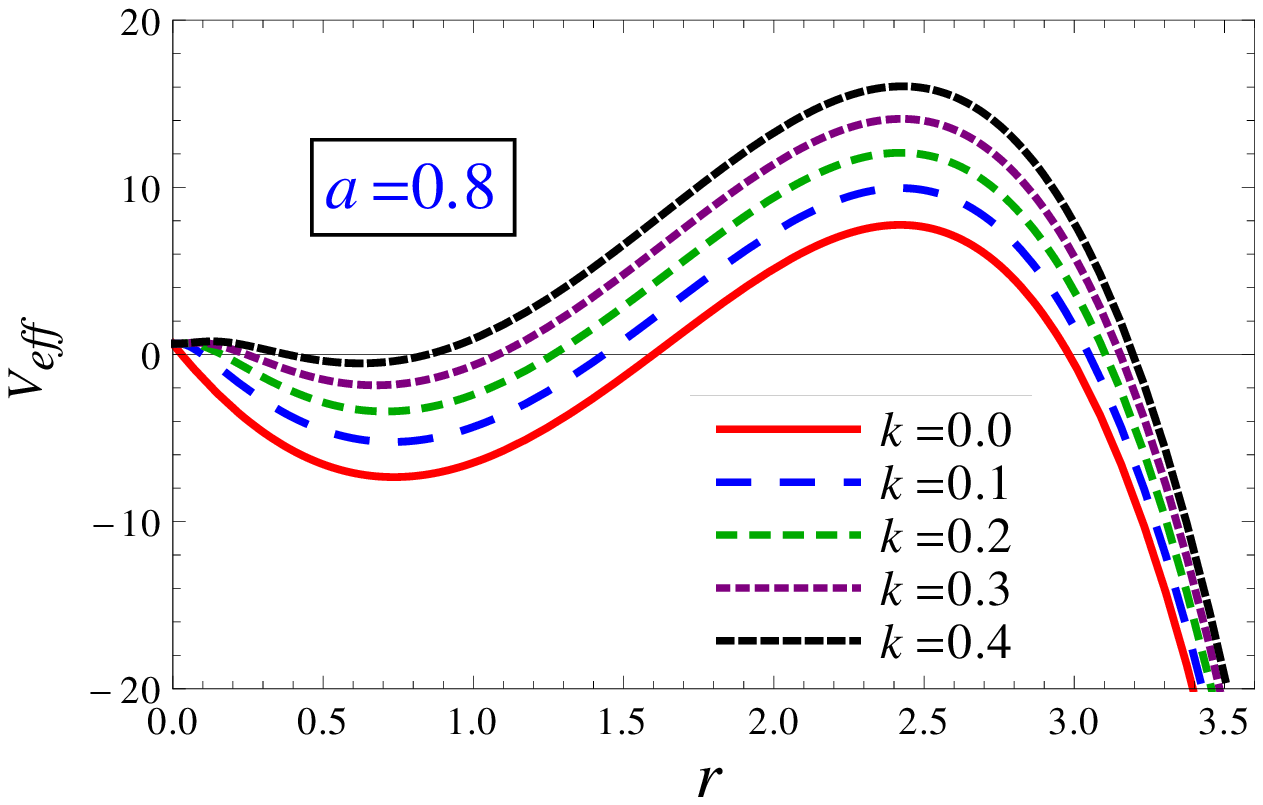}
	\end{tabular}
\caption{\label{eff} Plot showing the behavior of the effective potential for different values of $k$ with $\chi=4$, and $\zeta=1$.}
\end{figure*}

In this section, we present the calculation necessary for the shape of the black hole shadow, which demands the study of geodesics motion around the nonsingular black hole (\ref{metric}). The rotating nonsingular spacetime is characterized by three constants of the motion which are energy $E$, component of angular momentum $L_{z}$ and the Carter constant $\mathcal{K}$. To derive the geodesic equations, first we start with the Lagrangian
\begin{equation}\label{lagrangian}
\mathcal{L}=\frac{1}{2}g_{\mu \nu}\dot{x}^{\mu}\dot{x}^{\nu},
\end{equation}
where an overdot denotes partial derivative with respect to the affine parameter $\sigma$. The constants of motion $E$ and $L_{z}$ are given by \cite{Bardeen:1972fi}
\begin{equation}\label{E}
-E=\frac{\partial \mathcal{L}}{\partial \dot{t}}=g_{tt}\dot{t}+g_{t\phi}\dot{\phi},
\end{equation}
and
\begin{equation}\label{L}
L_{z}=\frac{\partial \mathcal{L}}{\partial \dot{\phi}}=g_{\phi \phi}\dot{\phi}+g_{t\phi}\dot{t},
\end{equation}
with $E$ and $L_z$, respectively, the energy and angular momentum. One can easily obtain the following geodesic equations by solving Eqs.~(\ref{E}) and (\ref{L}), simultaneously:
\begin{equation}\label{u^t}
\Sigma \frac{d t}{d \sigma} =  -a \left(aE \sin^2 \theta - L_{z}\right) + \frac{(r^2 + a^2) \left[(r^2+a^2)E-aL_{z}\right]}{r^2 + a^2- 2 M r e^{-k/r}},
\end{equation}
\begin{equation}\label{u^Phi}
\Sigma \frac{d \phi}{d \sigma} = -\left(aE - L_{z}\csc^2 \theta \right) + \frac{a \left[(r^2+a^2)E-aL_{z}\right]}{r^2 + a^2- 2 M r e^{-k/r}}.
\end{equation}
The remaining geodesic equations for the rotating nonsingular black hole with metric tensor $g^{\mu \nu}$ can be determined by using the Hamilton-Jacobi equation, which has the following form,
\begin{equation}\label{hje}
\frac{\partial S}{\partial \sigma} = -\frac{1}{2} g^{\mu\nu} \frac{\partial S}{\partial x^{\mu}} \frac{\partial S}{\partial x^{\nu}},
\end{equation}
where $S$ is the Jacobi action. The Hamilton-Jacobi Eq.~(\ref{hje}) has a solution of the following form,
\begin{equation}\label{hja}
S = \frac{1}{2} m_0^2 \sigma -Et + L_{z} \phi + S_{r}(r) + S_{\theta}(\theta),
\end{equation}
where $ m_0 $ corresponds to the rest mass of the particle, and $ S_{r} $ and $ S_{\theta} $ are functions of $ r $ and $ \theta $, respectively. Furthermore, one can obtain the geodesic equations by using Eqs.~(\ref{hje}) and (\ref{hja}) and separating out the coefficients of $r$ and $\theta$ equal to the Carter constant ($\pm \mathcal{K}$),
\begin{equation}\label{u^r}
\Sigma \frac{d r}{d \sigma} = \pm  \sqrt{\mathcal{R}},
\end{equation}
\begin{equation}\label{u^theta}
\Sigma \frac{d \theta}{d \sigma} = \pm  \sqrt{\Theta},
\end{equation}
where
\begin{eqnarray}
\label{R}
\mathcal{R} &=& \left[(r^2+a^2)E-aL_{z}\right]^2 \nonumber \\ 
&-&\left(r^2 + a^2- 2 M r e^{-k/r}\right) \left[m_0^2 r^2  +\mathcal{K}+ (L_{z}-a E)^2 \right], \nonumber \\ 
\end{eqnarray}
and
\begin{eqnarray}
\label{Th}
\Theta=\mathcal{K} +\cos^2 \theta \left(a^2E^2-L_{z}^2\csc^2 \theta \right),
\end{eqnarray}
where, without loss of generality one can take $m_{0}=1$ for particles, and $m_{0}=0$ for photons. We are interested in photon trajectories; therefore, the Eq.~(\ref{R}) transforms into
\begin{eqnarray}
\label{R1}
\mathcal{R}&=& \left[(r^2+a^2)E-aL_{z}\right]^2 \nonumber \\ 
&-&\left(r^{2}+a^{2}-2Mr e^{-k/r}\right) \left[\mathcal{K}+(L_{z}-a E)^2\right].
\end{eqnarray}
Now, we introduce dimensionless quantities such that $\chi=L_{z}/E$ and $\zeta=\mathcal{K}/E^2$, which are constant along the geodesics; therefore, in terms of these quantities Eq.~(\ref{R1}) looks like
\begin{eqnarray}
\label{R2}
\mathcal{R}&=& E^2\left[r^2+a^2-a\chi \right]^2 \nonumber \\ 
&-& E^2 \left(r^{2}+a^{2}-2Mr e^{-k/r}\right) \left[\zeta +(\chi -a )^2\right].\nonumber \\
\end{eqnarray}
Furthermore, to discuss the radial motion of the particle, the effective potential is a very important tool which can be calculated by using the equation \cite{Wei:2013kza}
\begin{eqnarray}
\left(\Sigma \frac{d r}{d \sigma}\right)^2 + V_{eff}=0,
\end{eqnarray}
hence, the expression of effective potential ($V_{eff}=-\mathcal{R}$) reads
\begin{eqnarray}\label{effpot}
V_{eff} &=& -E^2\left[r^2+a^2-a \chi \right]^2 \nonumber \\ 
&+& E^2 (r^2+a^2-2Mr e^{-k/r})\left[\zeta +(\chi -a)^2 \right].\nonumber \\ 
\end{eqnarray}
Indeed, the effective potential not only depends on parameter $k$ but also on $\chi$ and $\zeta$. Next, we are interested in circular orbits of the photons to find out $\chi$ and $\zeta$, which satisfy \cite{Bardeen:1972fi}
\begin{equation}\label{condition}
V_{eff} = 0 \;\;\;\;\; \text{and} \;\;\;\;\; \frac{d V_{eff}}{dr}=0.
\end{equation}
By using Eqs.~(\ref{effpot}) and (\ref{condition}), we can easily get the parameters $\chi$ and $\zeta$ in the following form:
\begin{eqnarray}\label{xi}
\chi &=& \frac{(r^2+a^2)r^2 e^{k/r}+M \left[(k-3 r) r^2+a^2 (k+r)\right]}{a M (k+r)-a r^2e^{k/r}},
\nonumber \\ 
\end{eqnarray}
\begin{eqnarray}\label{eta}
\zeta &=& -\frac{4 a^2Mr^4 (k-r) e^{k/r} +r^4(k M-3 Mr + r^2 e^{k/r})^2}{a^2 (k M+Mr-r^2 e^{k/r})^2}.
\nonumber \\ 
\end{eqnarray}
If one substitutes $k=0$, we get
\begin{eqnarray}\label{xi0}
\chi &=& \frac{(r^2+a^2)r^2 +M r(-3 r^2 +a^2 )}{a r(M-r)},
\end{eqnarray}
\begin{eqnarray}\label{eta0}
\zeta &=& \frac{4 a^2Mr^5 - r^5(r-3 M)^2}{a^2 r^2(M-r)^2},
\end{eqnarray}
which are exactly the same as the Kerr black hole \cite{Hioki:2009na}. Similarly, in the limit $k \rightarrow 0$, all quantities in this section correspond to the Kerr black hole \cite{Hioki:2009na}. The $V_{eff}$ dependence of the parameter $k$ with the particular choice of $\chi$ and $\zeta$ can be seen in Fig.~\ref{eff}, which shows that there are various bounds of the effective potential according to the parameter $k$.

\section{Nonsingular black hole shadow}
\label{shadow}
\begin{figure}[b]
	 \includegraphics[width=0.85\linewidth]{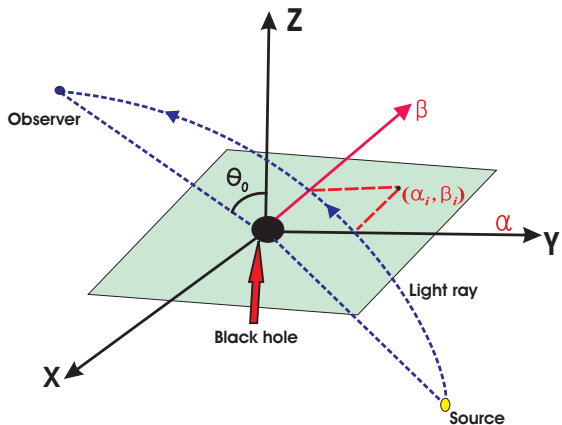}
	\caption{\label{celestial} Plot showing the geometrical representation of the celestial coordinates for a distant observer.}
\end{figure}

Next, we study the shadow of the rotating nonsingular black hole. We consider that the photons released from a bright object are coming towards the black hole, which is situated between an observer and a bright object. There should be  three possible trajectories of the photon geodesics around the black hole: (i) falling into the black hole (capture orbits), (ii) scattered away from the black hole to infinity (scatter orbits), and (iii) critical geodesics which separate the first two sets (unstable orbits). The observer which is far away from the black hole can see only the scattered photons while those photons which fall into the black hole form a dark region. This dark region is known as the black hole shadow. Furthermore, to study about this dark region or black hole shadow we are going to define the new coordinates ($\alpha,\beta$) \cite{Bardeen:1973gb}, which are known as celestial coordinates {(cf. Fig.~\ref{celestial})} defined by
\begin{figure}[t]
	 \includegraphics[scale=0.53]{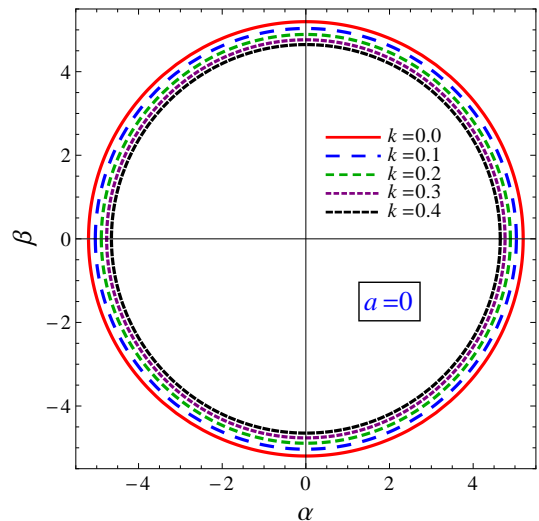}
	 \caption{\label{a=0} Plot showing the shapes of shadow cast by a nonrotating black hole ($a=0$) for different values of $k$.}
\end{figure}
\begin{figure}[!]
	 \includegraphics[scale=0.56]{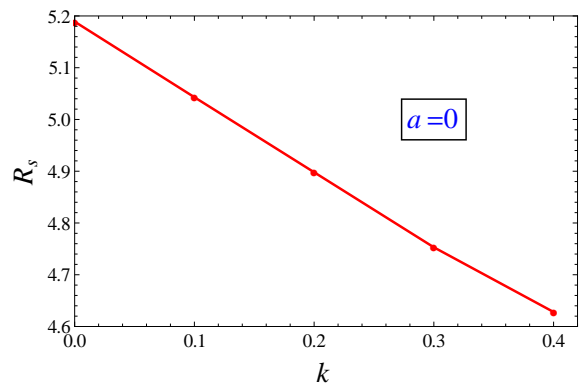}
	 \caption{\label{k-obs} Plot showing the behavior of the radius of the black hole shadow for different values of $k$.}
\end{figure}
\begin{figure*}
	\begin{tabular}{c c c}
 	\includegraphics[width=0.33\linewidth]{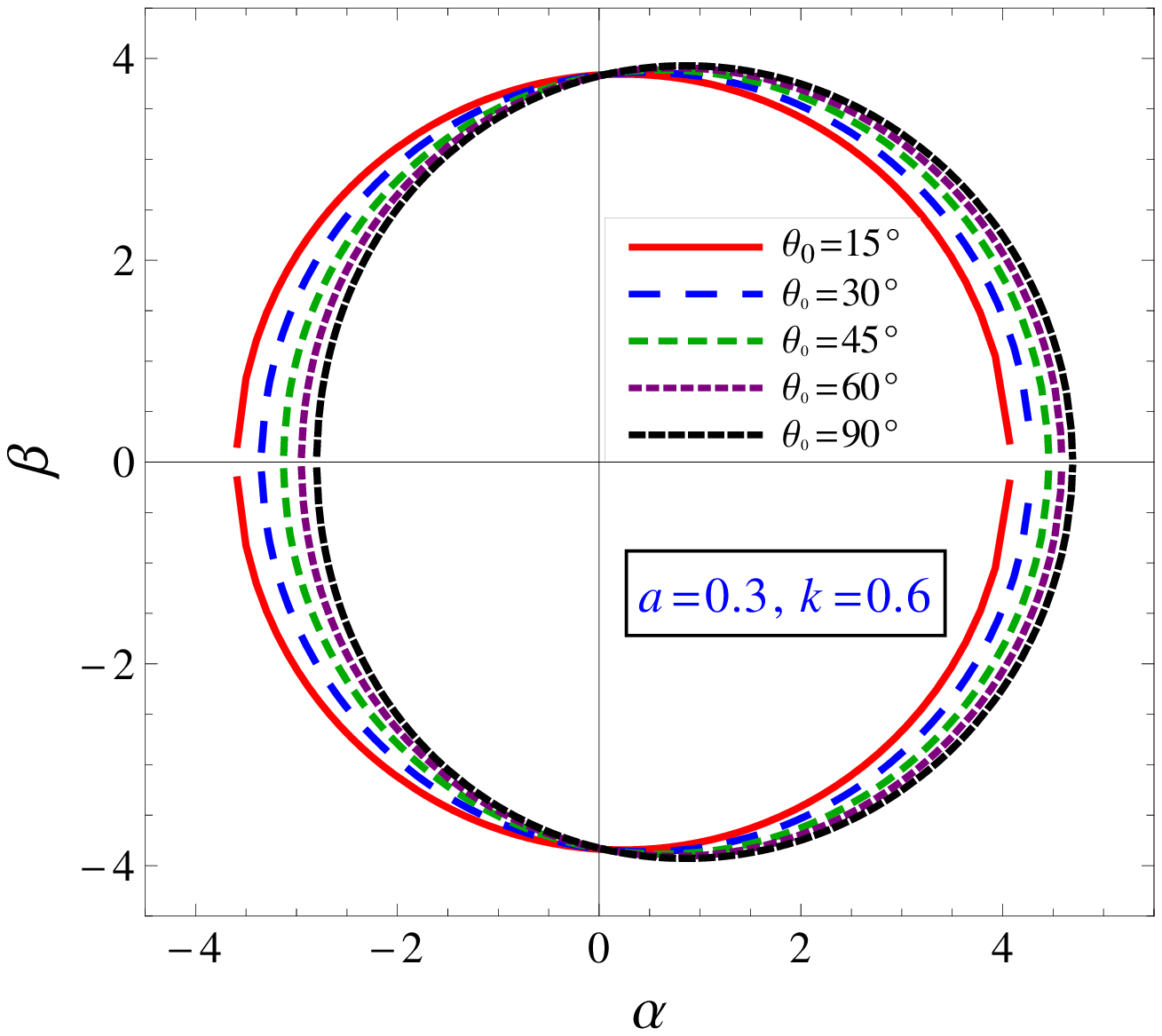}
 	\includegraphics[width=0.33\linewidth]{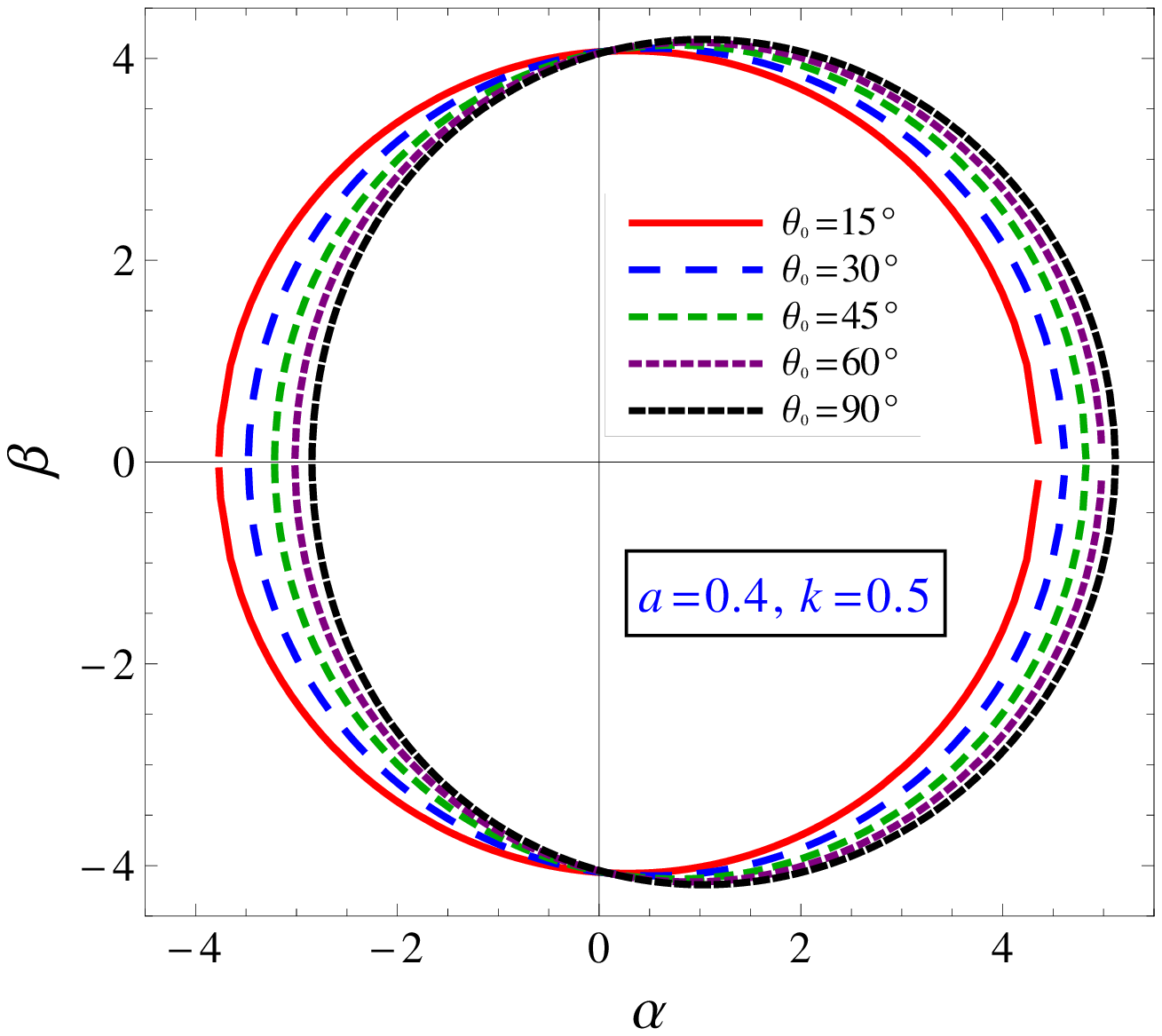}
 	\includegraphics[width=0.33\linewidth]{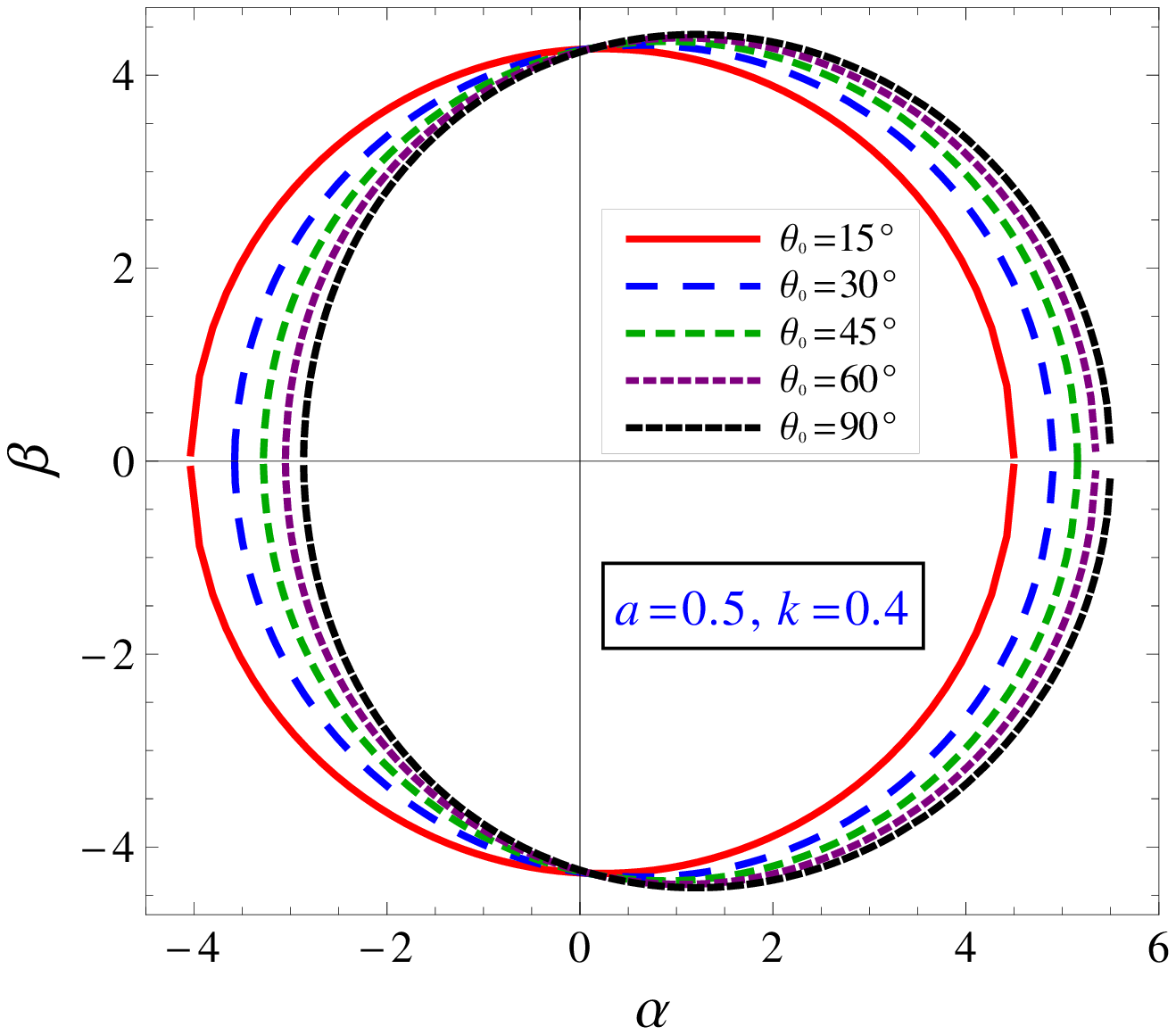}\\
 	\includegraphics[width=0.33\linewidth]{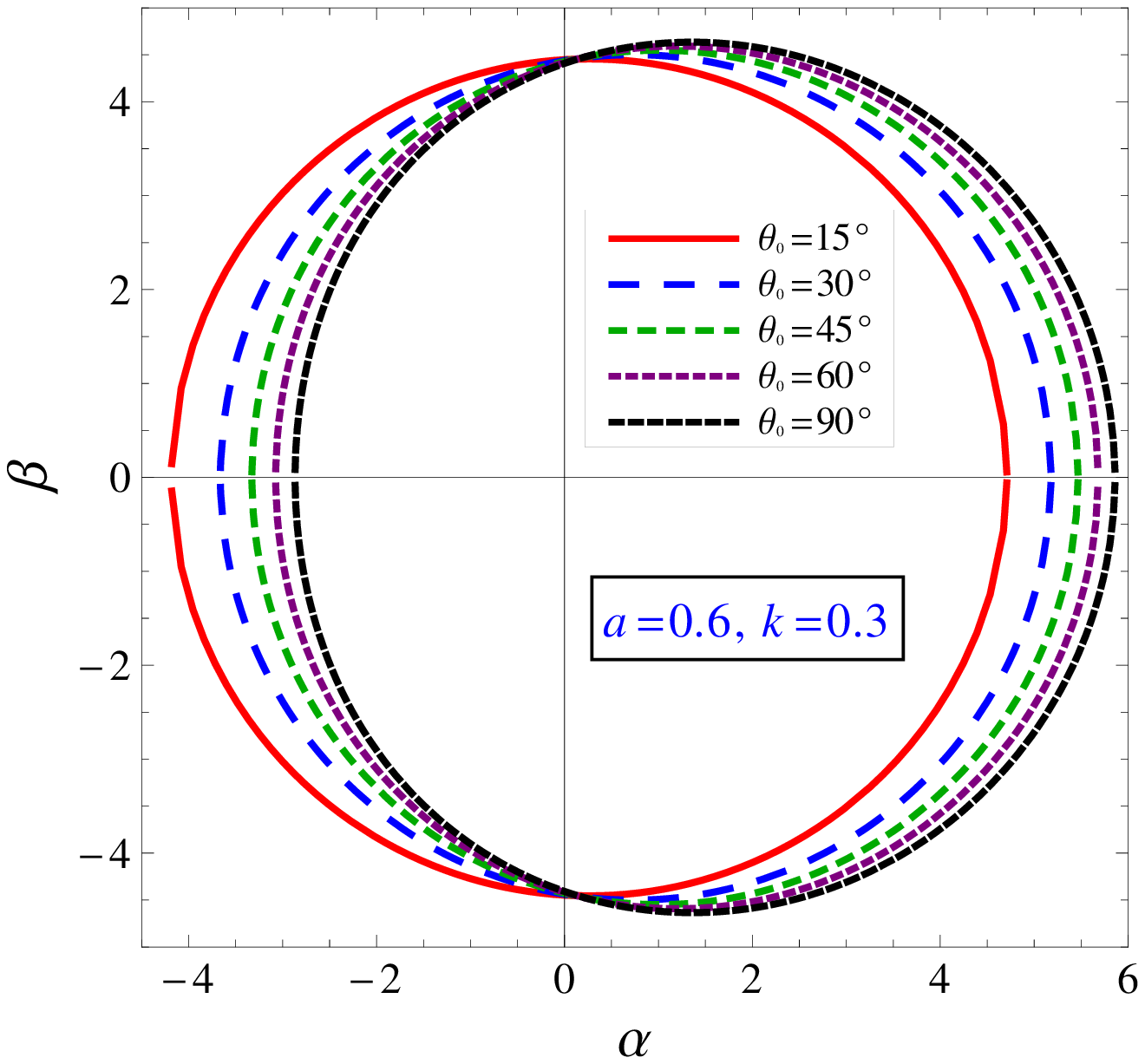}
 	\includegraphics[width=0.33\linewidth]{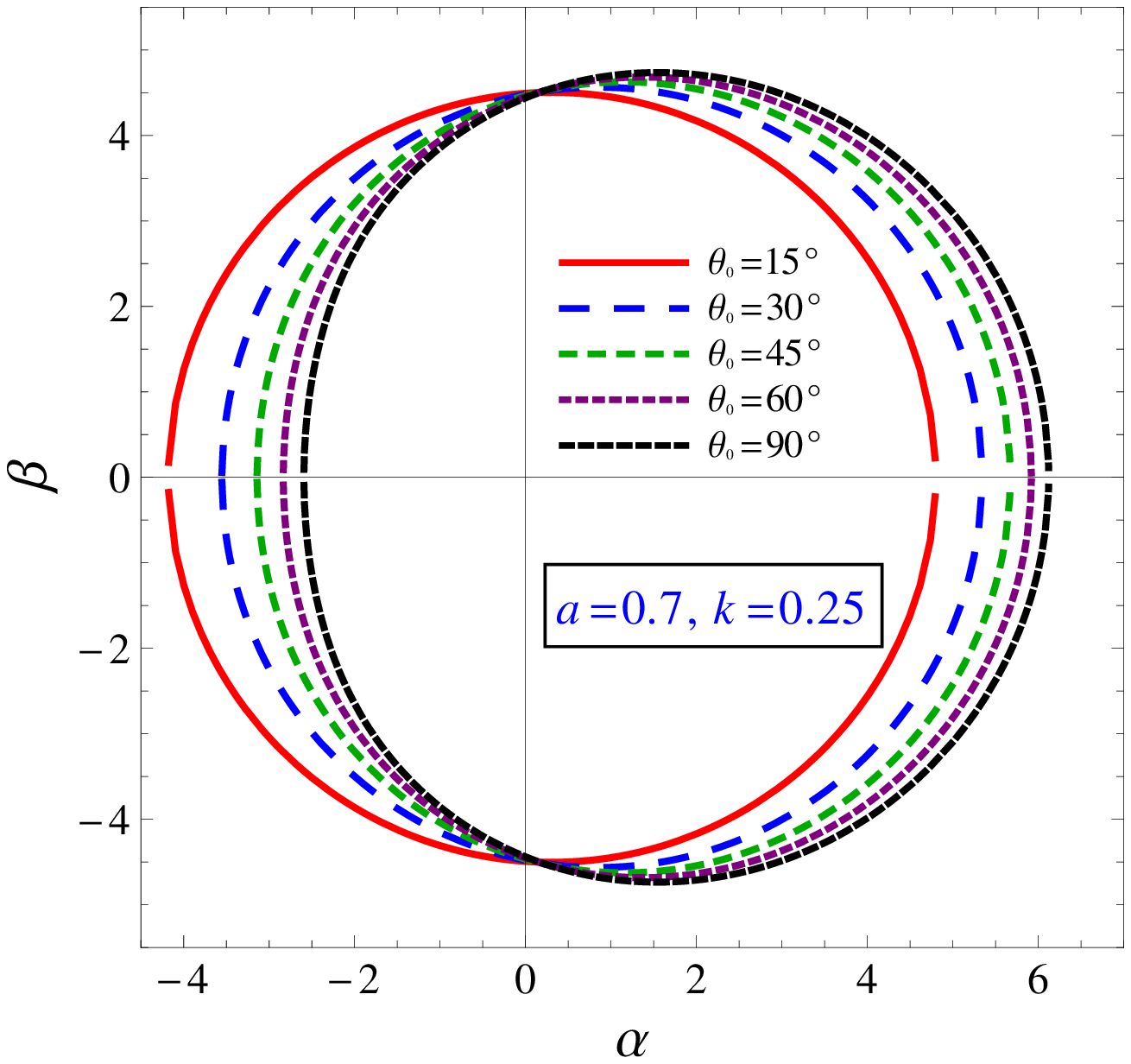}
 	\includegraphics[width=0.33\linewidth]{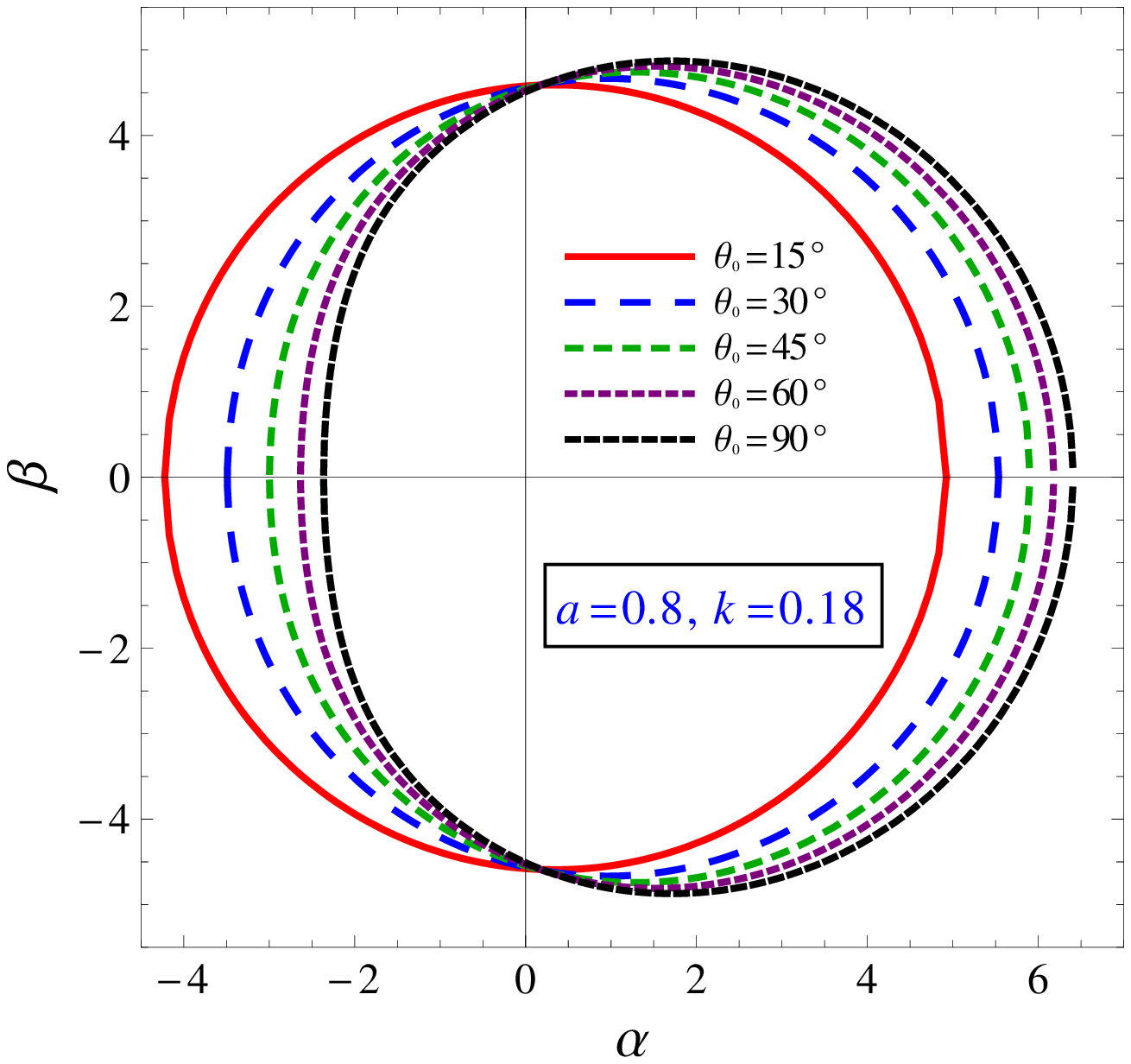}
	\end{tabular}
	\caption{\label{theta} Plot showing the shapes of the shadow cast by a rotating nonsingular black hole for some combinations of $a$ and $k$ with different values of inclination angle $\theta_{0}$.}
\end{figure*}
\begin{equation}
\label{alpha}
\alpha = \lim_{r_{0} \rightarrow \infty} \left(-r_{0}^2 \sin \theta_{0} \frac{d \phi}{dr} \right),
\end{equation}
\begin{equation}
\label{beta}
\beta = \lim_{r_{0} \rightarrow \infty} \left(r_{0}^2 \frac{d \theta}{dr} \right),
\end{equation}
where $r_{0}$ is the distance from the black hole to the observer and $\theta_{0}$ represents the inclination angle between the rotation axis of the black hole and the direction to an observer. By using Eqs.~(\ref{u^t}), (\ref{u^Phi}), (\ref{u^r}), and (\ref{u^theta}), we can easily calculate $d\phi /dr$ and $d \theta /dr$ when inserting these values in  Eqs.~(\ref{alpha}) and (\ref{beta}), where the celestial coordinates read \cite{Bardeen:1973gb,Chandrasekhar:1992}
\begin{equation}
\label{alpha1}
\alpha = -\chi \csc \theta_{0},
\end{equation}
\begin{equation}\label{beta1}
\beta = \pm \sqrt{\zeta +a^2 \cos^2 \theta_{0} -\chi^2 \cot^2 \theta_{0}}.
\end{equation}
where $\theta_{0}$ is the inclination angle between the rotation axis and the direction to an observer. If an observer is situated in the equatorial plane of the black hole, in this case the inclination angle is $\theta_{0}=\pi /2$. Hence, the Eqs.~(\ref{alpha1}) and (\ref{beta1}) transform into
\begin{equation}\label{alpha2}
\alpha = -\chi,
\end{equation}
\begin{equation}\label{beta2}
\beta = \pm \sqrt{\zeta}.
\end{equation}
Next, to visualize the shapes of the black hole shadow, we need to plot $\beta$ vs $\alpha$, where coordinates $\alpha$ and $\beta$ satisfy the following relation:
\begin{equation}\label{alpha&beta}
\alpha^2 + \beta^2 = \chi^2 +\zeta.
\end{equation}
Using Eqs.~(\ref{xi}), (\ref{eta}), and (\ref{alpha&beta}), we can obtain
\begin{eqnarray}\label{alpha&beta1}
\alpha^2 + \beta^2 &=& \frac{1}{(k M+ Mr- r^2 e^{k/r})^2}\Big[2 r^2 \nonumber \\
& \times& \left(r^4 e^{2 k/r}+M^2(k^2 -2 k r-3 r^2)\right) \nonumber \\
&+&a^2 \left(r^2 e^{k/r}+M(k +r) \right)^2\Big].
\end{eqnarray}
Now we can plot the shadow of the rotating nonsingular black hole (\ref{metric}) by using Eqs.~(\ref{alpha2}) and (\ref{beta2}). It is clear that the shapes or the contours of a black hole shadow depend on its spin, inclination angle and parameter $k$. We shall plot $\alpha$ vs $\beta$ to display the contour of the shadow of the nonsingular black hole for various values of $a$ and $k$ at different inclination angles.
\begin{figure*}
	\begin{tabular}{c c c}
 	\includegraphics[width=0.33\linewidth]{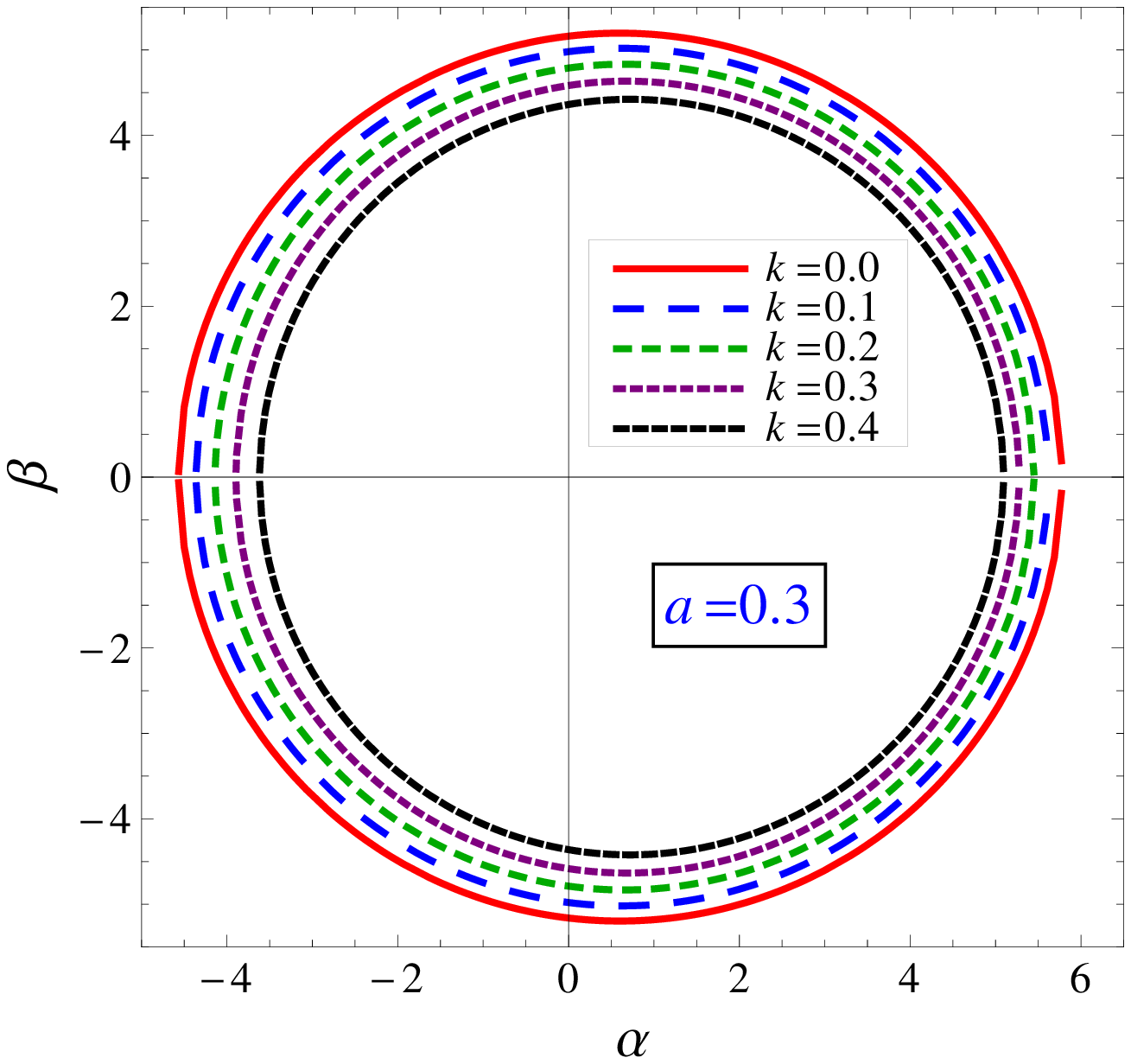}
 	\includegraphics[width=0.33\linewidth]{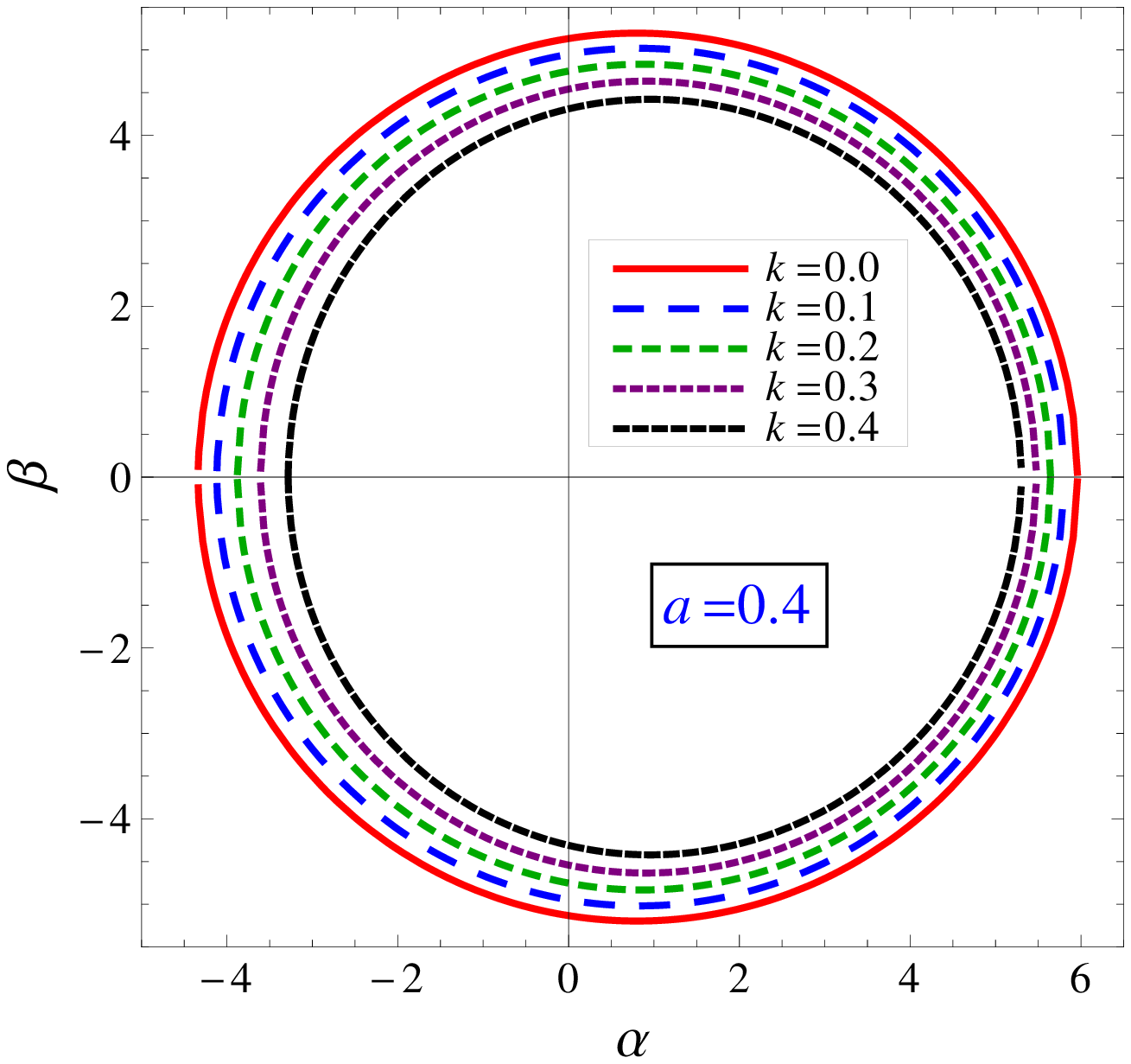}
 	\includegraphics[width=0.33\linewidth]{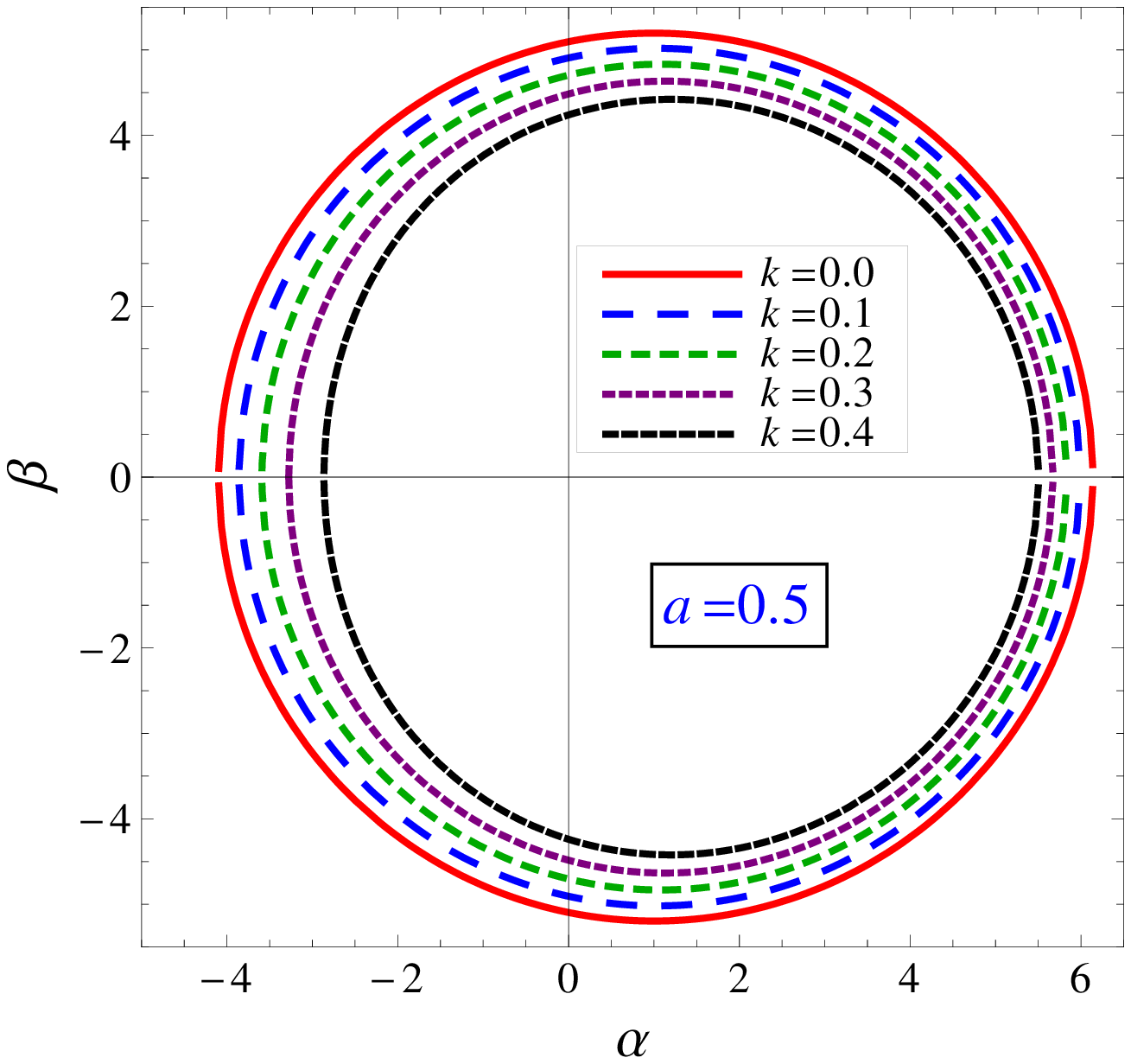}\\
 	\includegraphics[width=0.33\linewidth]{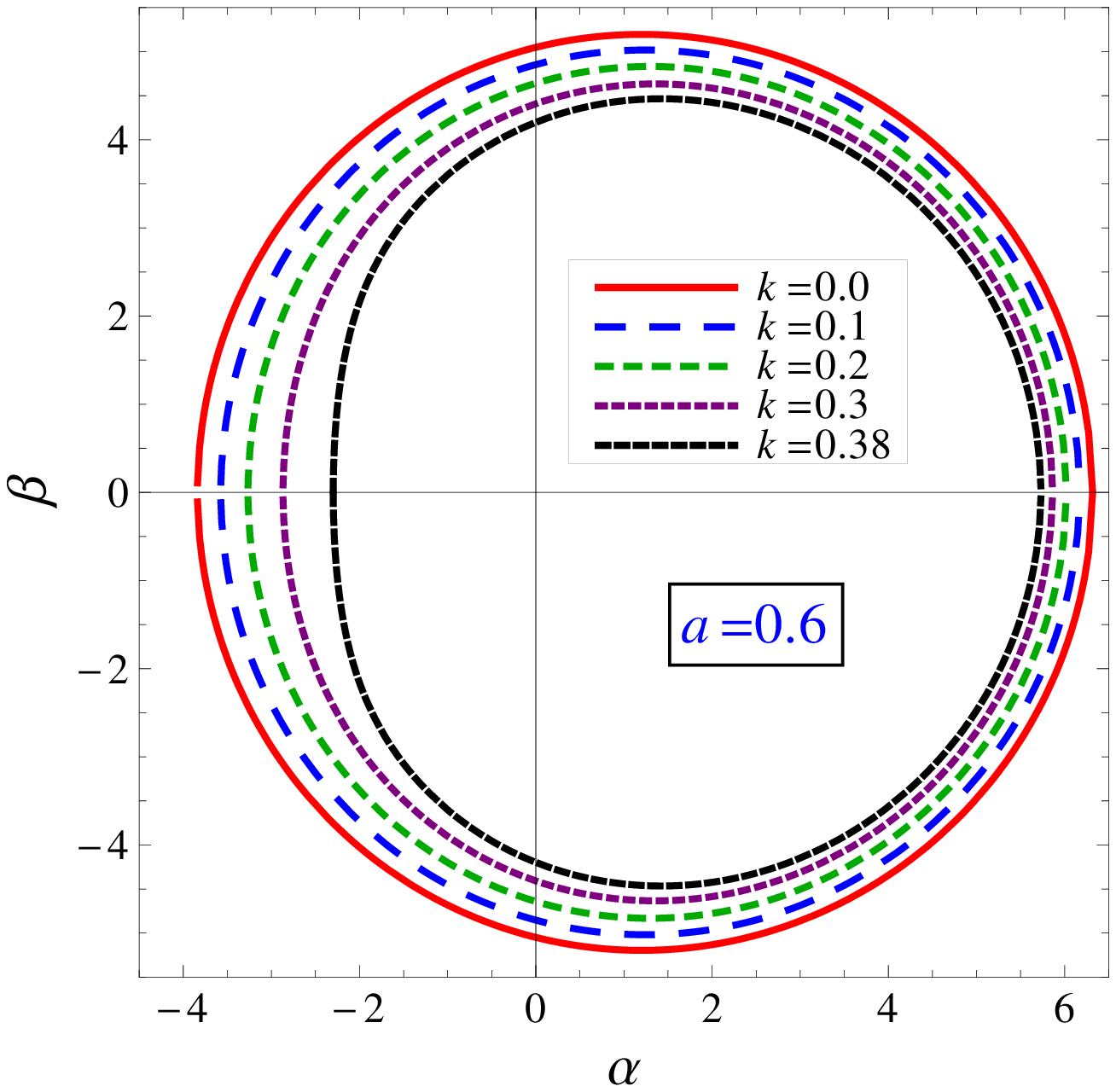}
 	\includegraphics[width=0.33\linewidth]{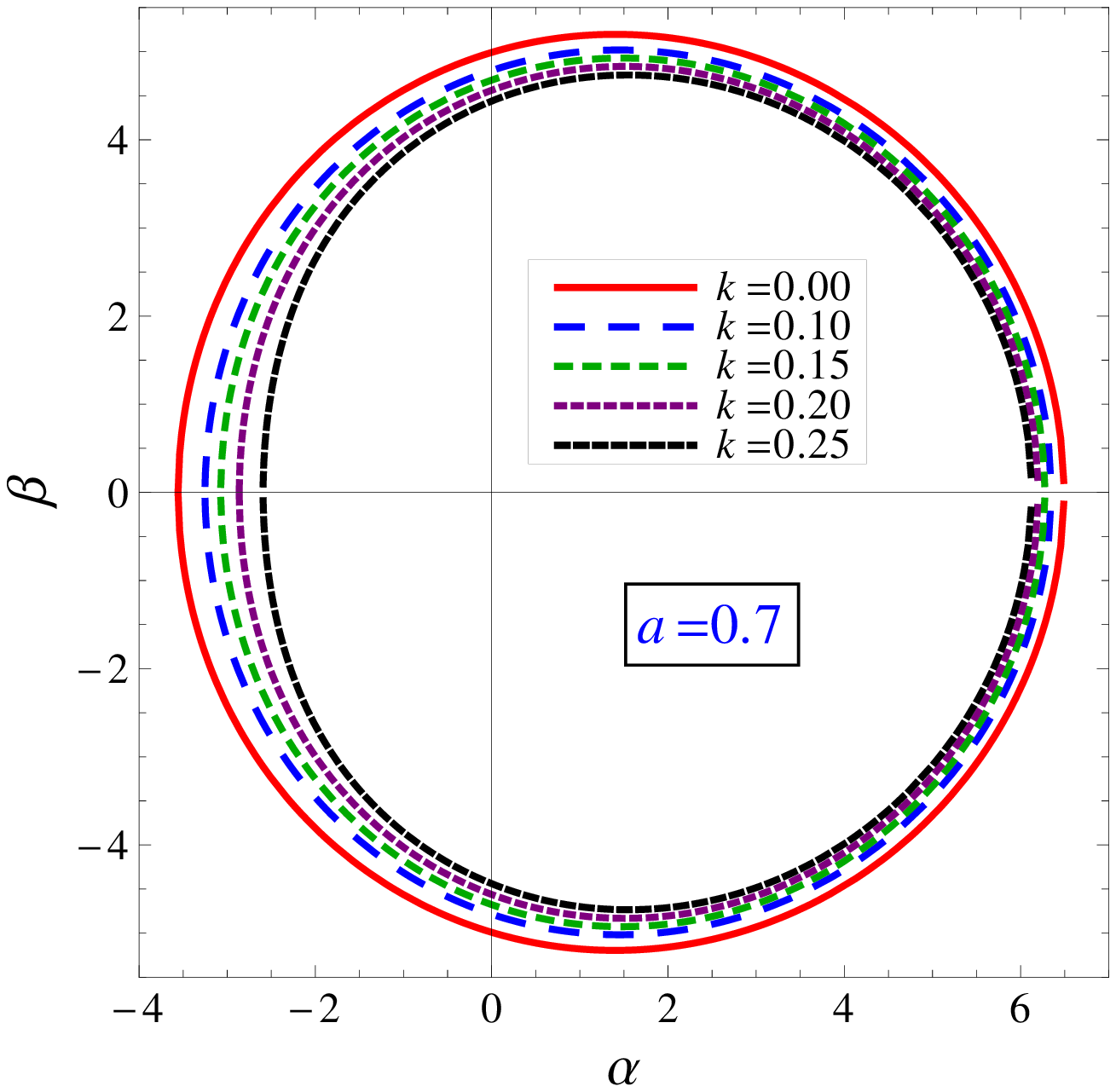}
 	\includegraphics[width=0.33\linewidth]{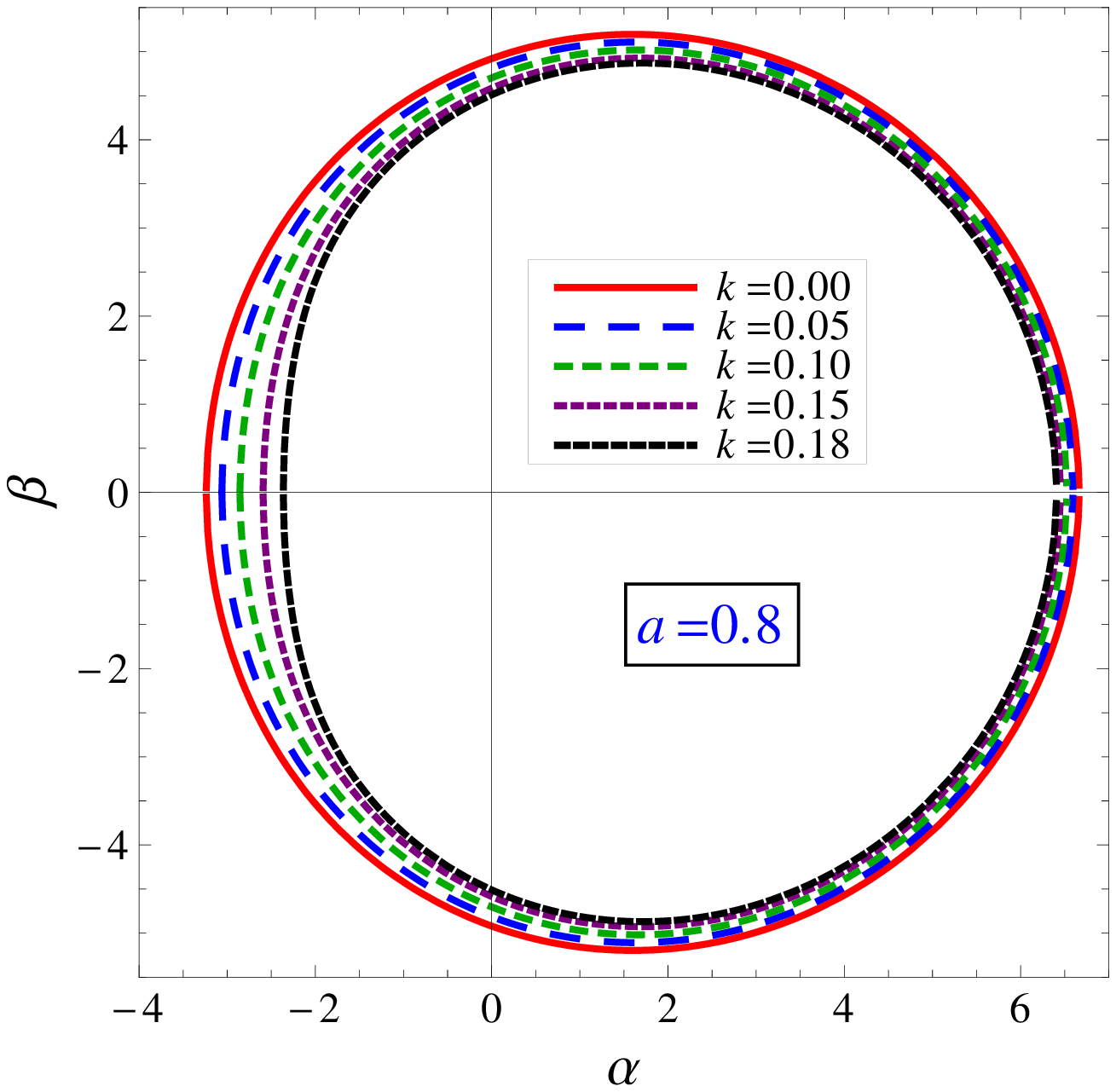}
	\end{tabular}
\caption{\label{k} Plot showing the shapes of the shadow cast by a rotating nonsingular black hole for the particular $a$ and different values of $k$.}
\end{figure*}

\subsection{Nonrotating case}
We first consider the nonrotating case, $a=0$, in which case the metric (\ref{metric}) reduces to one obtained in \cite{Culetu:2014lca}. In this case, the parameters $\alpha$ and $\beta$ satisfy
\begin{equation}\label{nonrot}
\alpha^2 + \beta^2 = \frac{2 r^2 (k^2-2 k r-3 r^2+ r^{4} e^{2 k/r})}{(r+ k- r^{2} e^{k/r})^2},
\end{equation}
where $r$ is the radius of circular orbits. The black hole shadow or plots of Eqs.~(\ref{alpha2}) and (\ref{beta2}) are presented in Fig.~\ref{a=0}. They are perfect circles. It can be {observed} from the plots that the parameter $k$ reduces the radius of the shadow (cf. Fig.~\ref{k-obs}). Consider the case when $k=0$, and Eq.~(\ref{nonrot}) transforms into
\begin{equation}\label{nonrot0}
\alpha^2 + \beta^2 = \frac{2 r^2( r^2-3)}{(1- r)^2}.
\end{equation}
Equation~(\ref{nonrot0}) is exactly the same as the one obtained for the Schwarzschild black hole \cite{Synge:1966}.
\begin{figure}
	 \includegraphics[width=0.8\linewidth]{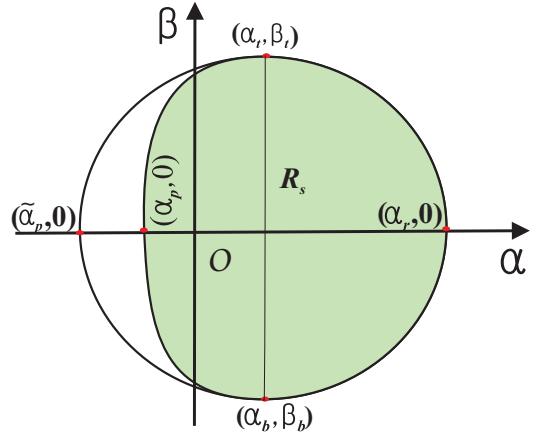}
	\caption{\label{observable} Plot showing the schematic representation of the observable.}
\end{figure} 
\begin{figure*}
 \includegraphics[width=0.48\linewidth]{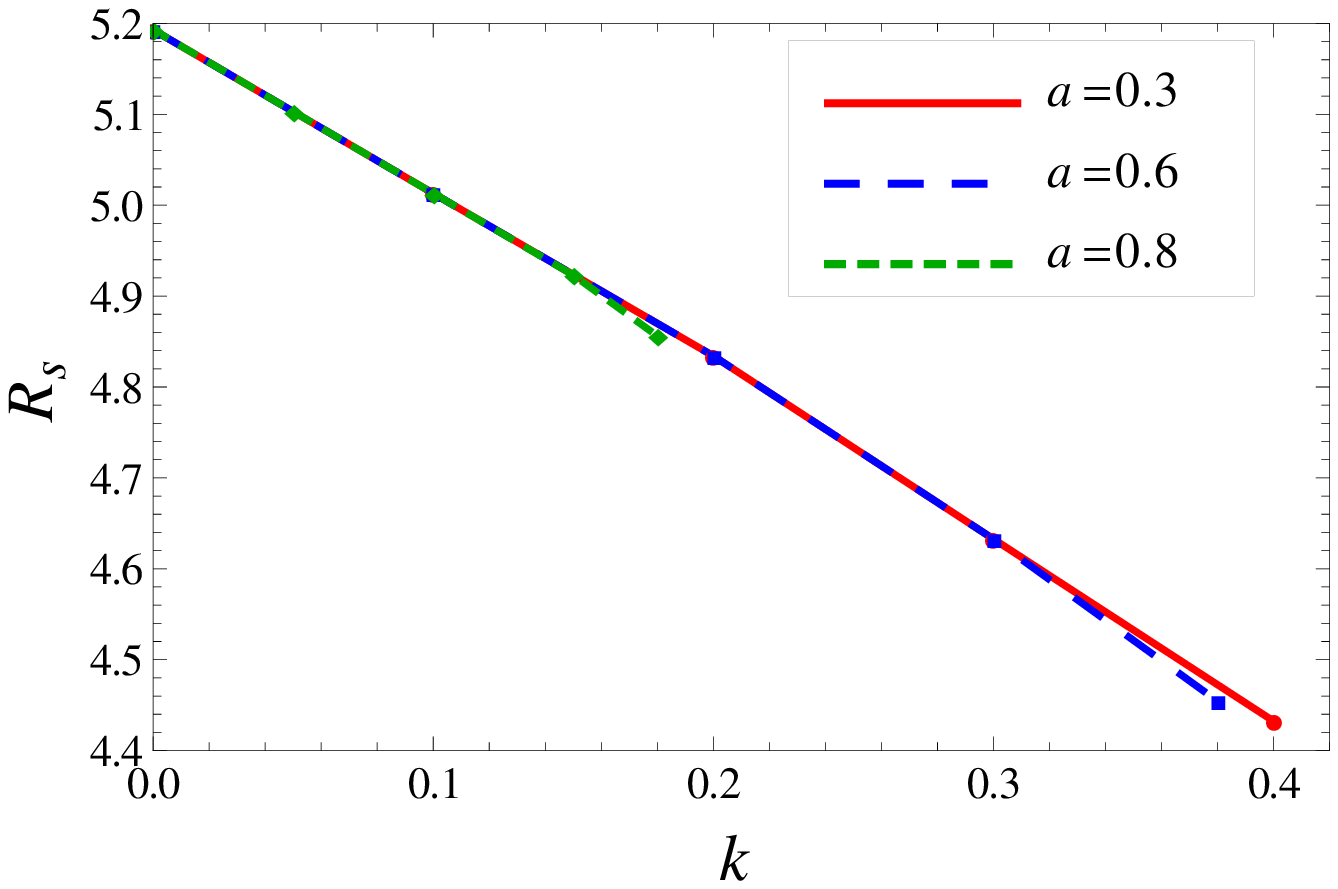}
 \includegraphics[width=0.48\linewidth]{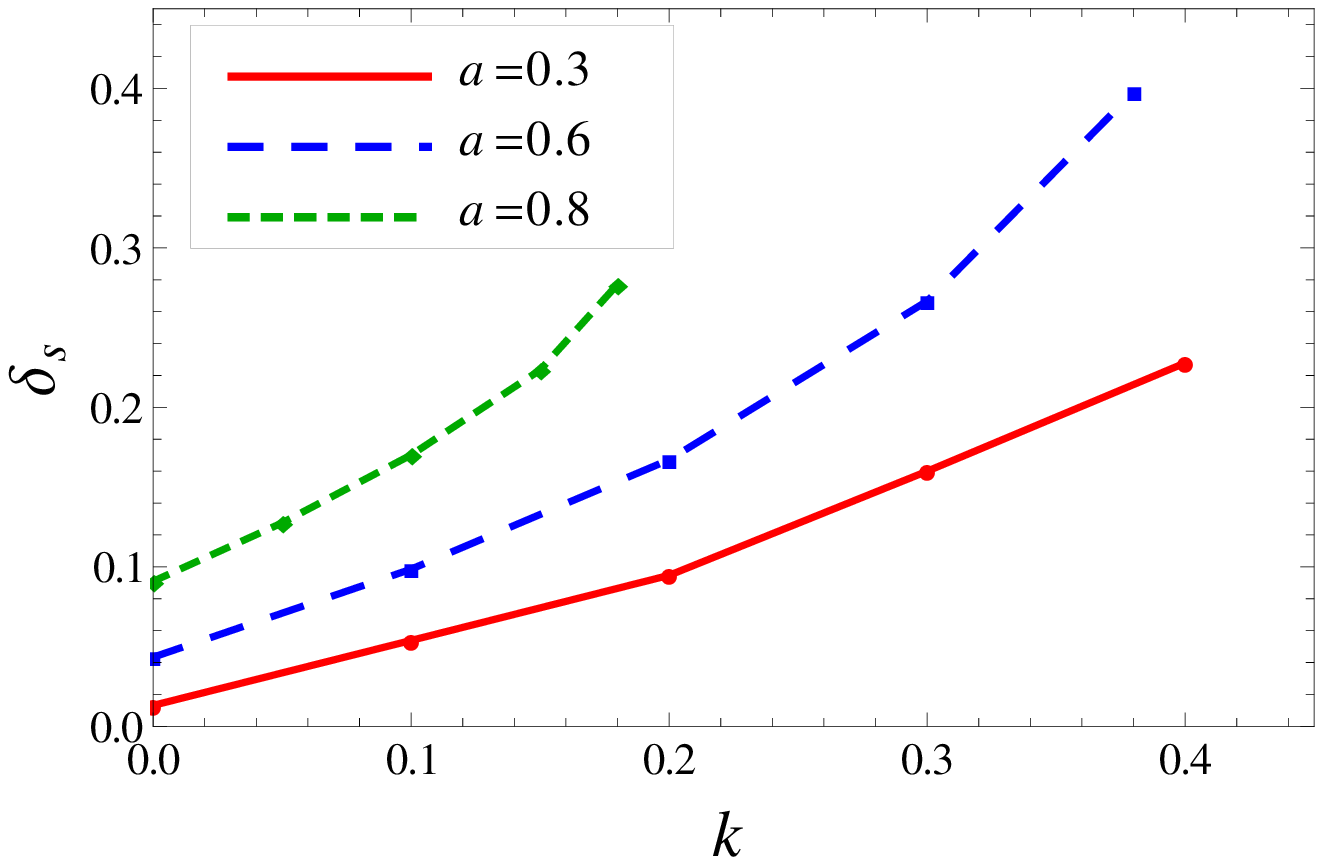}
\caption{\label{obs} Plot showing the behavior of the observable ($R_{s}$ and $\delta_{s}$) of shadow with $k$, for different values of $a$.}
\end{figure*}

\subsection{Rotating case}
It is shown that for a rotating black hole the shadow has an {atypical} deformation due to spin parameters instead of being a perfect circle. Here we wish to further analyze the effect of parameter $k$ on the apparent shapes of the rotating nonsingular black hole. Furthermore, if $k=0$, Eq.~(\ref{alpha&beta1}) reduces to
\begin{eqnarray}\label{alpha&beta01}
\alpha^2 + \beta^2 &=& \frac{2r^2(r^2 -3 M^2)+a^2 (r+M)^2}{(r-M)^2}.
\end{eqnarray}
Now, consider the case when the spin is nonzero, i.e., $a\neq0$. We have plotted the shapes of the shadow in Figs.~\ref{theta} and \ref{k}. In Fig.~\ref{theta}, we show the contours of the black hole shadow for various values of deviation parameter $k$ with different inclination angles $\theta_{0}$. Indeed, the shapes of the shadow are dependent on the inclination angle of the observer. Furthermore, Fig.~\ref{k} shows the effect of parameter $k$ on the shadow of a black hole. It is clear from Fig.~\ref{k} that when $a$ and $k$ take critical values, the shape of the shadow is more deformed.

Furthermore, to calculate the actual size and distortion of the black hole shadow, let us define the observable that characterizes the black hole shadow, namely, $R_{s}$ and $\delta_{s}$, where the parameters $R_{s}$ and  $\delta_{s}$ correspond to the actual size of the shadow and distortion in the shape of the shadow, respectively. The radius of the shadow can be calculated by considering a circle passing through the three points at the top, bottom, and rightmost corresponding to ($\alpha_{t}$,$\beta_{t}$), ($\alpha_{b}$,$\beta_{b}$), and ($\alpha_{r}$,$0$), respectively \cite{Hioki:2009na}. The definitions of these observables are given by \cite{Hioki:2009na}
\begin{equation}
\label{Rs}
R_{s}=\frac{(\alpha_{t}-\alpha_{r})^2+\beta_{t}^2}{2(\alpha_{t}-\alpha_{r})},
\end{equation}
and
\begin{equation}
\label{dels}
\delta_{s}=\frac{(\tilde{\alpha_{p}}-\alpha_{p})}{R_{s}},
\end{equation}
where $(\tilde{\alpha_{p}},0)$ and $(\alpha_{p},0)$ are the points where the reference circle and the contour of the shadow cut the horizontal axis at the opposite side of $(\alpha_{r},0)$ (cf., Fig.~\ref{observable}). The plots of the observable can be seen from Fig.~\ref{obs} which shows that when parameter $k$ increases, the radius of the shadow  decreases and the distortion increases.

\section{Energy emission rate}
\label{EE rate}
\begin{figure*}
 \includegraphics[width=0.48\linewidth]{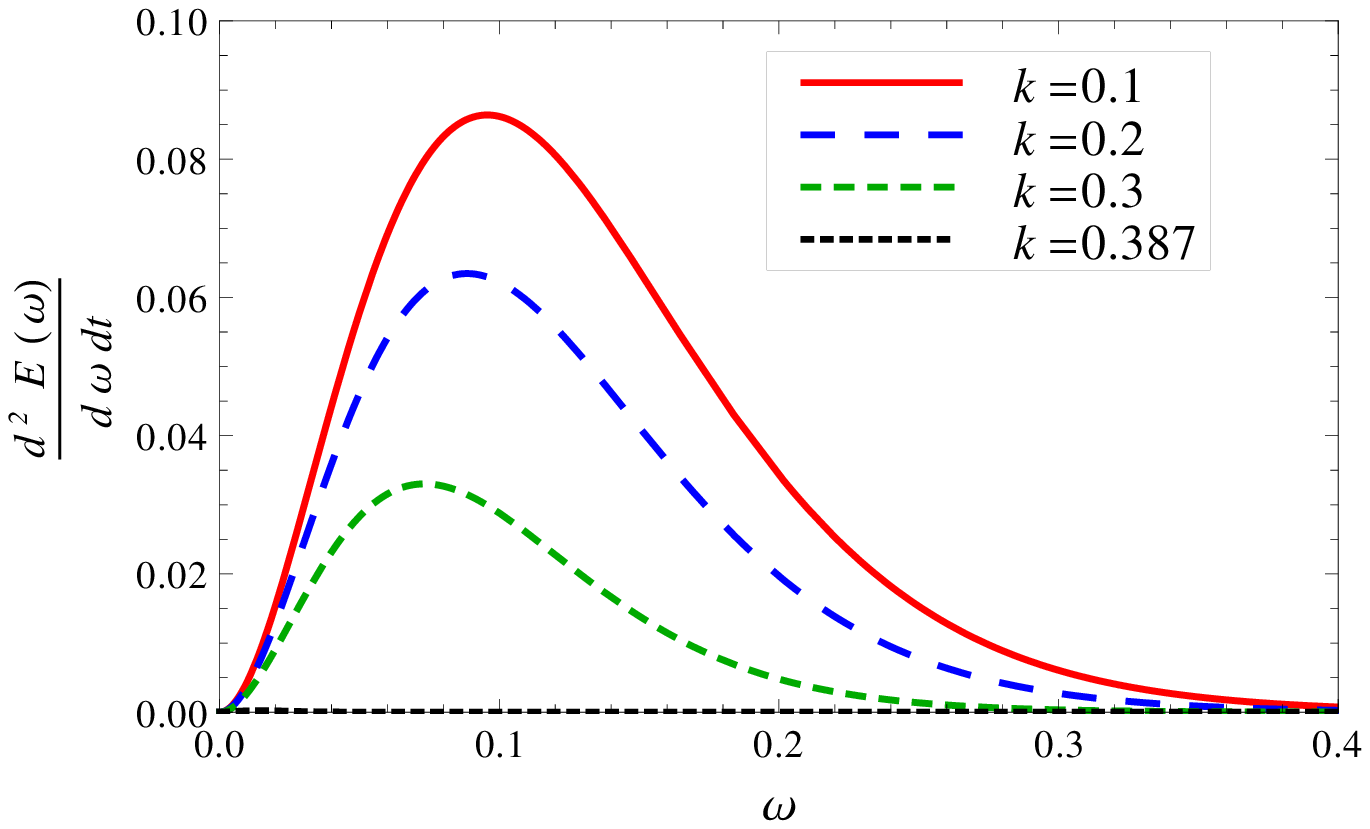}
 \includegraphics[width=0.48\linewidth]{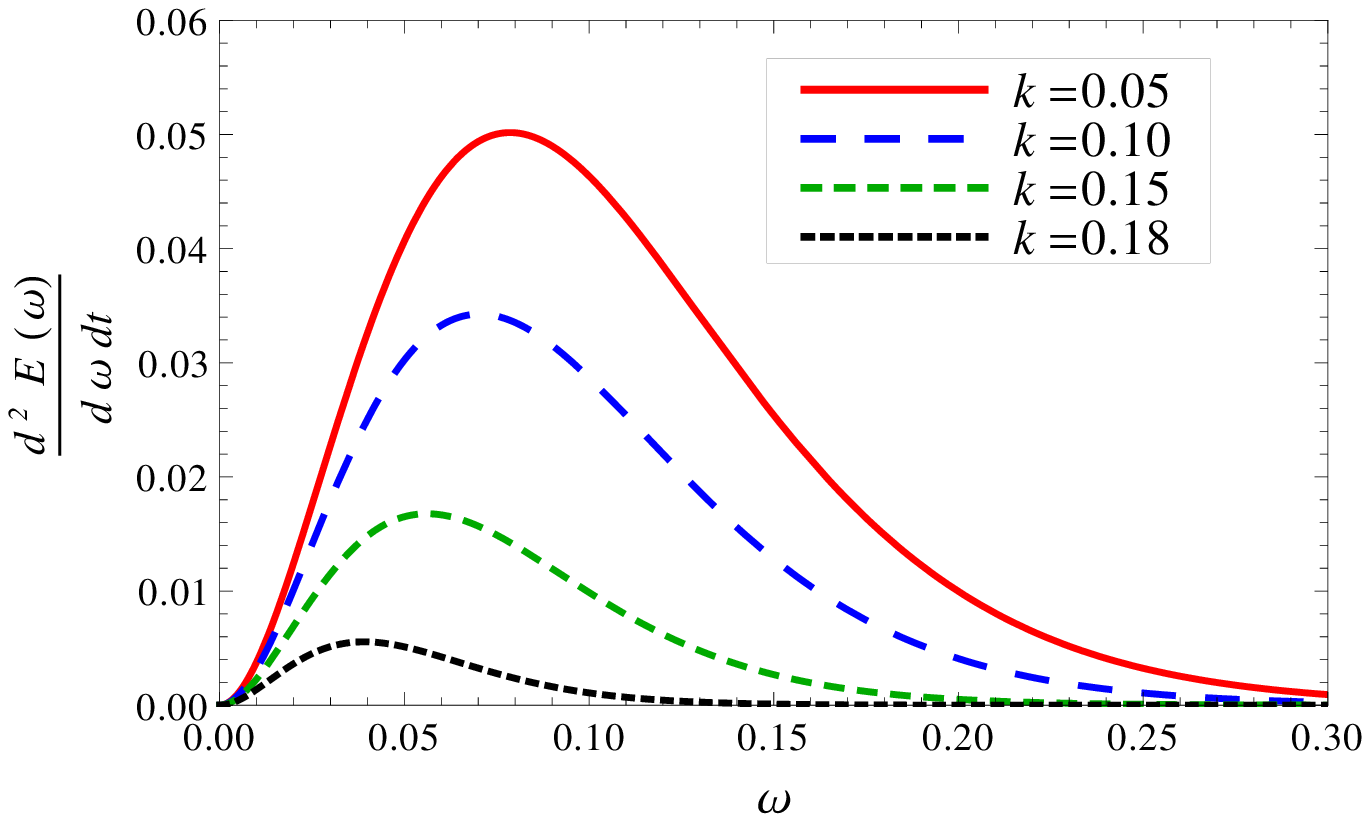}
\caption{\label{ee} Plot showing the behavior of the energy emission rate versus frequency with $k$. (Left) For $a=0.6$. (Right) For $a=0.8$.}
\end{figure*}

Next, we are interested in calculating the rate of energy emission by the rotating nonsingular black hole. First of all, we know that the limiting constant value of the absorption cross section for a spherically symmetric black hole is given by
\begin{equation}
\sigma_{lim} \approx \pi R_{s}^2.
\end{equation}
The black hole shadow is responsible for the high-energy absorption cross section for a distant observer. The  mentioned limiting constant value is derived in terms of geodesics and can be analyzed for the wave theories. For a black hole endowed with a photon sphere, the limiting constant value is the same as the geometrical cross section of this photon sphere. Now, by using this limiting constant value, we can compute the energy emission rate \cite{Wei:2013kza} by
\begin{equation}
\frac{d^2 E(\omega)}{d\omega dt}= \frac{2 \pi^2 R_{s}^2}{e^{\omega/ T}-1}\omega ^3,
\end{equation}
where
\begin{eqnarray}
T &=& \frac{1}{4 \pi r_{+}^2}\frac{r_{+}^2 (r_{+}-k)-a^2 (r_{+}+k)}{(r_{+}^2+a^2)}.
\end{eqnarray}
This generalizes the Hawking temperature of the Kerr black hole, and $r_{+}$ represents the event horizon. If $k=0$, then one obtains the Hawking temperature,
\begin{eqnarray}
T &=& \frac{1}{4 \pi r_{+}}\frac{r_{+}^2-a^2}{r_{+}^2+a^2},
\end{eqnarray}
for the Kerr black hole \cite{Good:2014uja}. The plots of the energy emission rate ($d^2 E(\omega)/d\omega dt$) versus the frequency ($\omega$) can be seen from Fig.~\ref{ee} for $a=0$ and $a=0.9$ by varying the parameter $k$. It is clear that, for small values of $k$, the energy emission  rate is high and if it takes a maximum value, then the energy emission rate is low.

\section{Conclusion}
\label{conclusion}
The investigation of the shadows of different types of black holes has been an important subject of research because that the observations of a black hole in the center of a galaxy may be determined \cite{Falcke:1999pj}. The shadow of a Kerr black hole is distorted mainly by its spin parameter and the inclination angle. It is a perfect circle for the Schwarzschild black hole.

In this paper, we make a qualitative analysis of the shapes of the black hole shadow cast by a rotating nonsingular black hole which has an additional deviation parameter $k$ and also perform a {detailed} study of its ergoregion. The dependence of  black hole shadows and ergoregions on spin $a$ and parameter $k$ are explicitly discussed. With the help of a study of null geodesics around the rotating nonsingular black hole, we construct the parameters $\zeta$ and $\chi$ by using the circular orbit conditions and discuss the behavior of the effective potential. We consider the black hole shadows for both nonrotating and rotating cases to bring out the effect of spin $a$ and parameter $k$. The shape of the shadows for the nonrotating black hole are perfect circles, and the size decreases with increasing value of $k$. In the next paragraph, we studied the shape of shadows for the rotating black hole. In this case, the shape is not a perfect circle, but it is deformed due to the presence of spin $a$ and parameter $k$. One can observe that the area of the ergoregion increases with an increase either in parameter $k$ or spin $a$. The observables $R_{s}$ and $\delta_{s}$ are defined to characterize the shape of the black hole shadow. We demonstrate the influence of the parameter $k$ on  observables $R_{s}$ and $\delta_{s}$. It shows that the value of $R_s$ decreases and $\delta_s$ increases with the parameter $k$.

The results obtained are similar to the Kerr black hole shadows \cite{Hioki:2009na}, but the size of the shadow decreases due to the deviation parameter $k$. The results obtained here may provide insight to into the qualitative features of a nonsingular black hole. We have done a detailed analysis of a rotating nonsingular black hole from the viewpoint of the ergoregion and shadow that contains the Kerr black hole analysis as the special case when the deviation parameter $k=0$.

\begin{acknowledgements}
M.A. acknowledges the University Grant Commission, India, for financial support through the Maulana Azad
National Fellowship For Minority Students scheme (Grant No.~F1-17.1/2012-13/MANF-2012-13-MUS-RAJ-8679). S.G.G. would like to thank SERB-DST for Research Project Grant No. SB/S2/HEP-008/2014. We also thank IUCAA for hospitality while a part of this work was done and ICTP for Grant No. OEA-NET-76.
\end{acknowledgements}


\begin{thebibliography}{00}

\bibitem{Tanaka:1995en}
  Y.~Tanaka {\it et al.},
  Nature (London) {\bf 375}, 659 (1995).

\bibitem{Fabian:1995qz}
  A.C.~Fabian, K.~Nandra, C.S.~Reynolds, W.N.~Brandt, C.~Otani, Y.~Tanaka, H.~Inoue, and K.~Iwasawa,
  Mon.\ Not.\ R.\ Astron.\ Soc.\  {\bf 277}, L11 (1995).

\bibitem{Genzel:2003as}
  R.~Genzel, R.~Schodel, T.~Ott, A.~Eckart, T.~Alexander, F.~Lacombe, D.~Rouan, and B.~Aschenbach,
  Nature (London) {\bf 425}, 934 (2003).

\bibitem{Connars:1980}
P.A.~Connors, R.F.~Stark, and T.~Piran, Astrophys.\ J.\ {\bf 235}, 224 (1980).

\bibitem{Asada:2003nf}
  H.~Asada, M.~Kasai, and T.~Yamamoto,
  Phys.\ Rev.\ D {\bf 67}, 043006 (2003).

\bibitem{Tanaka:1996ht}
  T.~Tanaka, Y.~Mino, M.~Sasaki, and M.~Shibata,
  Phys.\ Rev.\ D {\bf 54}, 3762 (1996).

\bibitem{Finn:2000sy}
  L.S.~Finn and K.~S.~Thorne,
  Phys.\ Rev.\ D {\bf 62}, 124021 (2000).

\bibitem{Takahashi:2004xh}
  R.~Takahashi,
  Astrophys.\ J.\  {\bf 611}, 996 (2004); 
  R.~Takahashi,
  Publ.\ Astron.\ Soc.\ Jpn.\  {\bf 57}, 273 (2005).

\bibitem{Bardeen:1973gb}
  J.M.~Bardeen, 
  in {\it Black holes, in Proceeding of the Les Houches Summer School, Session} 215239, edited by C. De Witt and B.S. De Witt and B.S. De Witt (Gordon and Breach, New York, 1973).

\bibitem{Chandrasekhar:1992}
 S.~Chandrasekhar,
 \textit{The Mathematical Theory of Black Holes} (Oxford University Press, New York, 1992).

\bibitem{Falcke:1999pj}
  H.~Falcke, F.~Melia and E.~Agol,
  Astrophys.\ J.\  {\bf 528}, L13 (2000).

\bibitem{Bozza:2009yw}
  V.~Bozza,
  Gen.\ Relativ.\ Gravit.\  {\bf 42}, 2269 (2010).

\bibitem{Holz:2002uf}
  D.E.~Holz and J.A.~Wheeler,
  Astrophys.\ J.\  {\bf 578}, 330 (2002).

\bibitem{Synge:1966}
  J.L.~Synge,
  Mon.\ Not.\ R.\ Astron.\ Soc.\  {\bf 131}, 463 (1966).

\bibitem{Luminet:1979}
   J.P.~Luminet
  Astron.\ Astrophys. {\bf 75}, 228 (1979).

\bibitem{Virbhadra:2008ws}
  K.S.~Virbhadra,
  Phys.\ Rev.\ D {\bf 79}, 083004 (2009).

\bibitem{Bisnovatyi-Kogan:2015dxa}
  G.S.~Bisnovatyi-Kogan and O.Y.~Tsupko,
  Plasma\ Phys.\ Rep.\  {\bf 41}, 562 (2015).

\bibitem{Morozova:2013dxb}
V.S. Morozova, B.J. Ahmedov, and A.A. Tursunov,
Astrophys.\ Space\ Sci.\ {\bf 346}, 513 (2013).

\bibitem{Nedkova:2013msa}
  P.G.~Nedkova, V.K.~Tinchev, and S.S.~Yazadjiev,
  Phys.\ Rev.\ D {\bf 88}, 124019 (2013).

\bibitem{Zakharov:2005ek}
  A.F.~Zakharov, F.~De Paolis, G.~Ingrosso, and A.A.~Nucita,
  New Astron.\  {\bf 10}, 479 (2005);
   A.F.~Zakharov, F.~De Paolis, G.~Ingrosso, and A.A.~Nucita,
  Astron.\ Astrophys.\ {\bf 442}, 795 (2005);
  F.~De Paolis, G.~Ingrosso, A.A.~Nucita, A.~Qadir and A.F.~Zakharov,
  Gen.\ Relativ.\ Gravit.\  {\bf 43}, 977 (2011).

\bibitem{Bambi:2008jg}
  C.~Bambi and K.~Freese,
  Phys.\ Rev.\ D {\bf 79}, 043002 (2009).

\bibitem{Bozza:2007gt}
  V.~Bozza and G.~Scarpetta,
  Phys.\ Rev.\ D {\bf 76}, 083008 (2007).

\bibitem{Hioki:2009na}
  K.~Hioki and K.~i.~Maeda,
  Phys.\ Rev.\ D {\bf 80}, 024042 (2009).

\bibitem{Takahashi:2005hy}
  R.~Takahashi,
  Publ.\ Astron.\ Soc.\ Jap.\  {\bf 57}, 273 (2005).
  
\bibitem{Wei:2013kza}
  S.~W.~Wei and Y.~X.~Liu,
  J.\ Cosmol.\ Astropart.\ Phys.\ 11 (2013) 063.
  
\bibitem{Abdujabbarov:2012bn}
  A.~Abdujabbarov, F.~Atamurotov, Y.~Kucukakca, B.~Ahmedov, and U.~Camci,
  Astrophys.\ Space Sci.\  {\bf 344}, 429 (2013).

\bibitem{Amarilla:2011fx}
  L.~Amarilla and E.~F.~Eiroa,
  Phys.\ Rev.\ D {\bf 85}, 064019 (2012).

\bibitem{Amarilla:2013sj}
  L.~Amarilla and E.~F.~Eiroa,
  Phys.\ Rev.\ D {\bf 87}, 044057 (2013).

\bibitem{Atamurotov:2013sca}
  F.~Atamurotov, A.~Abdujabbarov, and B.~Ahmedov,
  Phys.\ Rev.\ D {\bf 88}, 064004 (2013).

\bibitem{Grenzebach:2014fha}
  A.~Grenzebach, V.~Perlick, and C.~L{\"a}mmerzahl,
  Phys.\ Rev.\ D {\bf 89}, 124004 (2014).

\bibitem{Cunha:2015yba} 
  P.V.~P.~Cunha, C.A.~R.~Herdeiro, E.~Radu, and H.F.~Runarsson,
  Phys.\ Rev.\ Lett.\  {\bf 115}, 211102 (2015).

\bibitem{Abdujabbarov:2015xqa} 
  A.A.~Abdujabbarov, L.~Rezzolla, and B.~J.~Ahmedov,
  Mon.\ Not.\ R.\ Astron.\ Soc.\  {\bf 454}, 2423 (2015).
  
\bibitem{Papnoi:2014aaa}
  U.~Papnoi, F.~Atamurotov, S.~G.~Ghosh, and B.~Ahmedov,
  Phys.\ Rev.\ D {\bf 90}, 024073 (2014).

\bibitem{Stuchlik:2014qja}
  Z.~Stuchlík and J.~Schee,
  Int.\ J.\ Mod.\ Phys.\ D {\bf 24}, 1550020 (2015).

\bibitem{Bambi:2014nta}
  C.~Bambi,
  Phys.\ Lett.\ B {\bf 730}, 59 (2014).

\bibitem{Ghosh:2014mea}
  S.G.~Ghosh, P.~Sheoran, and M.~Amir,
  Phys.\ Rev.\ D {\bf 90}(10), 103006 (2014).

\bibitem{Schee:2015nua}
  J.~Schee and Z.~Stuchlik,
  J.\ Cosmol.\ Astropart.\ Phys.\ 06 (2015) 048.

\bibitem{Amir:2015pja}
  M.~Amir and S.G.~Ghosh,
  J.\ High\ Energy\ Phys.\ 07 (2015) 015.

\bibitem{Ghosh:2015pra}
  S.~G.~Ghosh and M.~Amir,
  Eur.\ Phys.\ J.\ C {\bf 75}, 553 (2015).

\bibitem{Li:2013jra}
  Z.~Li and C.~Bambi,
  J.\ Cosmol.\ Astropart.\ Phys.\  01 (2014) 041.

\bibitem{Tinchev:2015apf}
  V.K.~Tinchev,
  Chin.\ J.\ Phys.\  {\bf 53}, 110113 (2015).

\bibitem{Abdujabbarov:2016hnw} 
  A.~Abdujabbarov, M.~Amir, B.~Ahmedov, and S.G.~Ghosh,
  Phys.\ Rev.\ D {\bf 93}, 104004 (2016).

\bibitem{Ansoldi:2008jw}
  S.~Ansoldi,
  arXiv:0802.0330.

\bibitem{Bambi:2013ufa}
  C.~Bambi and L.~Modesto,
  Phys.\ Lett.\ B {\bf 721}, 329 (2013).

\bibitem{Toshmatov:2014nya}
  B.~Toshmatov, B.~Ahmedov, A.~Abdujabbarov, and Z.~Stuchlik,
  Phys.\ Rev.\ D {\bf 89}, 104017 (2014).

\bibitem{Ghosh:2014hea}
  S.G.~Ghosh and S.D.~Maharaj,
  Eur.\ Phys.\ J.\ C {\bf 75}, 7 (2015).

\bibitem{Bardeen} J.M.~Bardeen, in \textit{Conference Proceedings of GR5, Tbilisi}, USSR, 1968, p. 174.

\bibitem{ABG99} E.~Ay{\'o}n-Beato, A.~Garc{\'i}a, Phys. Lett. B {\bf 493}, 149 (2000).

\bibitem{AGB} E.~Ay{\'o}n-Beato, A.~Garc{\'i}a, Phys. Rev. Lett. {\b 80},  5056 (1998).

\bibitem{AGB1} E.~Ay{\'o}n-Beato and A.~Garc{\'i}a, Gen.\ Rel.\ Grav.\  {\bf 31}, 629 (1999).

\bibitem{AGB2} E.~Ay{\'o}n-Beato and A.~Garc{\'i}a, Gen.\ Relativ.\ Gravit.\ {\bf 37}, 635 (2005).

\bibitem{regular} I.~Dymnikova, Gen.\ Relativ.\ Gravit.\ {\bf 24}, 235 (1992).

\bibitem{regular1} I.~Dymnikova, Classical\ Quantum\ Gravity\ {\bf 21}, 4417  (2004).

\bibitem{regular2} K.A.~Bronnikov, Phys.\ Rev.\ D\ {\bf 63}, 044005 (2001).

\bibitem{regular3} S.~Shankaranarayanan and N.~Dadhich, Int.\ J.\ Mod.\ Phys.\ D {\bf 13}, 1095 (2004).

\bibitem{Hayward} S.A.~Hayward, Phys.\ Rev.\ Lett.\  {\bf 96}, 031103 (2006).

\bibitem{Culetu:2014lca} 
  H.~Culetu,
  Int.\ J.\ Theor.\ Phys.\  {\bf 54}, 2855 (2015).

\bibitem{lbev} L.~Balart and E.C.~Vagenas,  Phys.\ Lett.\ B {\bf 730}, 14 (2014).

\bibitem{Balart:2014cga} L.~Balart and E.C.~Vagenas, Phys.\ Rev.\ D {\bf 90}(12), 124045 (2014).

\bibitem{Xiang} L.~Xiang, Y.~Ling, and Y.G.~Shen, Int.\ J.\ Mod.\ Phys.\ D {\bf 22}, 1342016 (2013).

\bibitem{Ghosh:2014pba}
  S.G.~Ghosh,
  Eur.\ Phys.\ J.\ C {\bf 75}, 532 (2015).

\bibitem{Kerr:1963ud}
  R.P.~Kerr,
  Phys.\ Rev.\ Lett.\  {\bf 11}, 237 (1963).

\bibitem{schw}
 K.~Schwarzschild, 
 Uber das Gravitationsfeld eines Manpunktes nach der Einsteinschen Theorie, Sitzber. Deutsch. Akad. Wiss. 
 Berlin, Kl. Math. Phys. Tech., 189, (1916).

\bibitem{Newman:1965my}
  E.T.~Newman, R.~Couch, K.~Chinnapared, A.~Exton, A.~Prakash, and R.~Torrence,
  J.\ Math.\ Phys.\  {\bf 6}, 918 (1965). 
  
\bibitem{Bambi:2011mj} 
  C.~Bambi,
  Mod.\ Phys.\ Lett.\ A {\bf 26}, 2453 (2011); 
  Astronomy\ Review\  {\bf 8}, 4 (2013).

\bibitem{Bambi:2013sha} 
  C.~Bambi,
  J.\ Cosmol.\ Astropart.\ Phys.\ 08 (2013) 055.

\bibitem{Gou:2011nq} 
  L.~Gou {\it et al.},
  Astrophys.\ J.\  {\bf 742}, 85 (2011).

\bibitem{Gou:2013dna} 
  L.~Gou {\it et al.},
  Astrophys.\ J.\  {\bf 790}, 29 (2014).
   
\bibitem{Penrose:1971uk} 
  R.~Penrose and R.M.~Floyd,
  Nature (London) {\bf 229}, 177 (1971).
  
\bibitem{Bardeen:1972fi}
  J.M.~Bardeen, W.H.~Press and S.A.~Teukolsky,
  Astrophys.\ J.\  {\bf 178}, 347 (1972).

\bibitem{Good:2014uja} 
  M.R.R.~Good and Y.C.~Ong,
  Phys.\ Rev.\ D {\bf 91}, 044031 (2015).
\end{thebibliography}
\end{document}